\documentclass[fleqn,usenatbib]{mnras}

\usepackage{newtxtext,newtxmath}
\usepackage{textcomp}

\usepackage[T1]{fontenc}

\DeclareRobustCommand{\VAN}[3]{#2}
\let\VANthebibliography\thebibliography
\def\thebibliography{\DeclareRobustCommand{\VAN}[3]{##3}\VANthebibliography}

\usepackage{graphicx}	

\title[Structure, stability and optical photoabsorption properties of small Ti$_n$C$_x$ clusters]{Structure, stability and optical absorption spectra of small Ti$_n$C$_x$ clusters: a first-principles approach}

\author[Sergio G\'amez-Valenzuela, {\it et al.}]{
Sergio G\'amez-Valenzuela,$^{1}$
Julio A. Alonso,$^{2}$
Gonzalo Santoro$^{3}$
and Jos\'e I. Mart\'{\i}nez$^{3}$\thanks{E-mail: joseignacio.martinez@icmm.csic.es}
\\
$^{1}$Department of Physical Chemistry, University of Malaga, Campus de Teatinos s/n, ES-29071, Malaga, Spain\\
$^{2}$Department of Theoretical and Atomic Physics and Optics, University of Valladolid, ES-47011 Valladolid, Spain\\
$^{3}$ESISNA Group, Institute of Materials Science of Madrid (ICMM-CSIC), ES-28049 Madrid, Spain
}

\date{Accepted XXX. Received YYY; in original form ZZZ}

\pubyear{2021}

\begin{document}
\label{firstpage}
\pagerange{\pageref{firstpage}--\pageref{lastpage}}
\maketitle

\begin{abstract}
Titanium-carbide molecular clusters are thought to form in the circumstellar envelopes (CSEs) of carbon-rich Asymptotic Giant Branch stars (AGBs) but, to date, their detection has remained elusive. To facilitate the astrophysical identification of those clusters in AGBs and post-AGBs environments, the molecular structures and optical absorption spectra of small Ti$_n$C$_x$ clusters, with n = 1--4 and x = 1--4, and some selected larger clusters, Ti$_3$C$_8$, Ti$_4$C$_8$, Ti$_6$C$_{13}$, Ti$_7$C$_{13}$, Ti$_8$C$_{12}$, Ti$_9$C$_{15}$, and Ti$_{13}$C$_{22}$, have been calculated. The density functional formalism, within the B3LYP approximation for electronic exchange and correlation, was used to find the lowest energy structures. Except the clusters having a single Ti atom, the rest exhibit three-dimensional structures. Those are formed by a Ti fragment surrounded in general by carbon dimers. The optical spectra of Ti$_n$C$_x$, computed by time-dependent density functional theory, using the corrected CAM-B3LYP functional, show absorption features in the visible and near infrared regions which may help in the identification of these clusters in space. In addition, most of the clusters have sizable electric dipole moments, allowing their detection by radioastronomical observations.
\end{abstract}




\section{Introduction} \label{intro}

Titanium carbide (TiC) dust is predicted to form in the inner regions of the circumstellar envelopes (CSEs) of carbon rich (C-rich) Asymptotic Giant Branch stars (AGBs)~\citep{ref1}. Due to its high refractory nature, TiC is considered to be the first condensate in the atmosphere of C-rich AGBs and may act as seed for the heterogeneous nucleation of dust. The identification of TiC crystals embedded in graphite spherules extracted from the Murchinson meteorite supports this prediction~\citep{ref2, ref3} but, to date, the detection of molecules comprised of titanium and carbon that may serve as precursors for the formation of TiC dust has remained elusive.

In the inner regions of the CSE of AGBs, titanium is in atomic form presenting fairly low abundances. The reaction of Ti with acetylene (C$_2$H$_2$) has been suggested as a possible pathway for the formation of stoichiometric TiC molecules~\citep{ref4, ref5}, and chemical equilibrium calculations of AGBs atmospheres show that the depletion of atomic Ti starts at stellar radii in which most carbon is locked into C$_2$H$_2$. However, the predicted abundance of the diatomic molecule TiC is too low and the high stability of large clusters such as Ti$_8$C$_{12}$, which might be formed through the reaction of smaller Ti$_n$C$_x$ clusters, suggests that it is a reasonable candidate as molecular precursor of TiC dust. Indeed, chemical equilibrium calculations show that Ti$_8$C$_{12}$ replaces atomic Ti as the main Ti reservoir at 2--3 stellar radii~\citep{ref1}.
        
\begin{figure*}
\includegraphics[width=\textwidth]{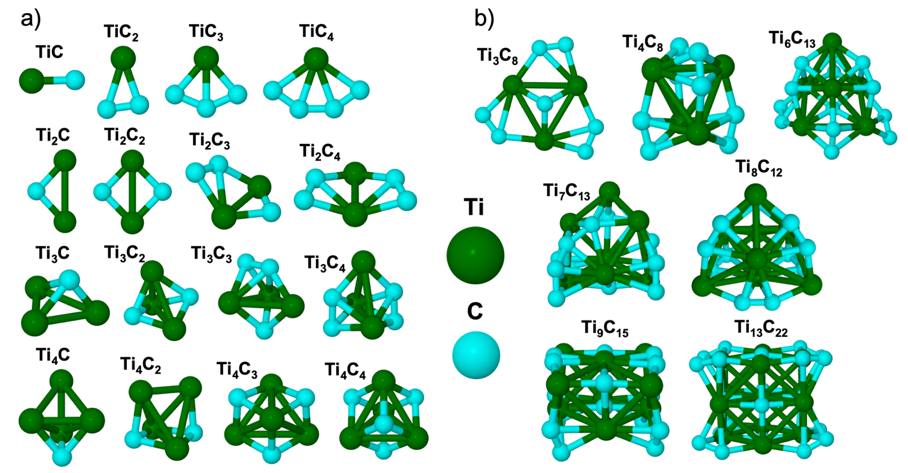}
\caption{Lowest energy structures of Ti$_n$C$_x$ clusters: (a) with n = 1--4 and x = 1--4, and (b) some selected larger clusters: Ti$_3$C$_8$, Ti$_4$C$_8$, Ti$_6$C$_{13}$, Ti$7$C$_{13}$, Ti$_8$C$_{12}$, Ti$_9$C$_{15}$, and Ti$_{13}$C$_{22}$. Large (green) and small (cyan) spheres represent Ti and C atoms, respectively.}
\label{Fig1}
\end{figure*}
    
On the other hand, TiC nanocrystals were suggested as carriers of the 21 $\mu$m emission feature in C-rich protoplanetary nebulae (PPNe) ~\citep{ref6}. However, the low abundance of Ti in these environments is not consistent with this carrier~\citep{ref7, ref8} and the possibility of a hydrocarbon-based carrier has been considered instead, although coated TiC grains could circumvent the constrains imposed by the Ti abundance in post-AGB environments~\citep{ref9}.   

The solid-state phase diagram of the Ti-C alloy~\citep{ref10} shows a very stable compound centered at Ti$_{55}$C$_{45}$ with a wide composition range. This compound has a high melting temperature of 3066$^\circ$C. These features reveal that the formation of strong Ti-C bonds is favorable, and this is also the case for Ti--C nanoclusters, which exhibit highly stable structures at particular stoichiometries. 

Three decades ago, by reacting metal vapors with hydrocarbons, Castleman and coworkers discovered a class of highly stable cage-like clusters with formula M$_8$C$_{12}$, the metallocarbohedrenes, or metcars~\citep{ref11,ref12,ref13,ref14}, where M denotes an early transition metal (mainly, Ti, V, Zr, Nb, Hf, Mo, Cr, and Fe). Later on, ~\cite{ref15} performed similar experiments by reacting laser-vaporized Ti with CH$_{4}$, and observed the formation of Ti$_{8}$C$_{12}$ along with several other clusters including Ti$_n$C$_{2n}$ stoichiometries. The mass spectrometric studies of anionic Ti$_n$C$_x$$^{-}$ clusters performed by~\cite{ref16,ref17} revealed new stable anionic clusters and evidenced the importance of C$_2$ dimers in stabilizing the structure of these clusters. These authors measured the photoelectron spectra of the most stable anionic clusters, namely: Ti$_3$C$_8$$^{-}$, Ti$_4$C$_8$$^{-}$, Ti$_6$C$_{13}$$^{-}$, Ti$_7$C$_{13}$$^{-}$, Ti$_9$C$_{15}$$^{-}$, Ti$_{13}$C$_{22}$$^{-}$; and also of clusters with a single Ti atom: TiC$_x$$^{-}$, x = 2--5.

\begin{figure}
\includegraphics[width=\columnwidth]{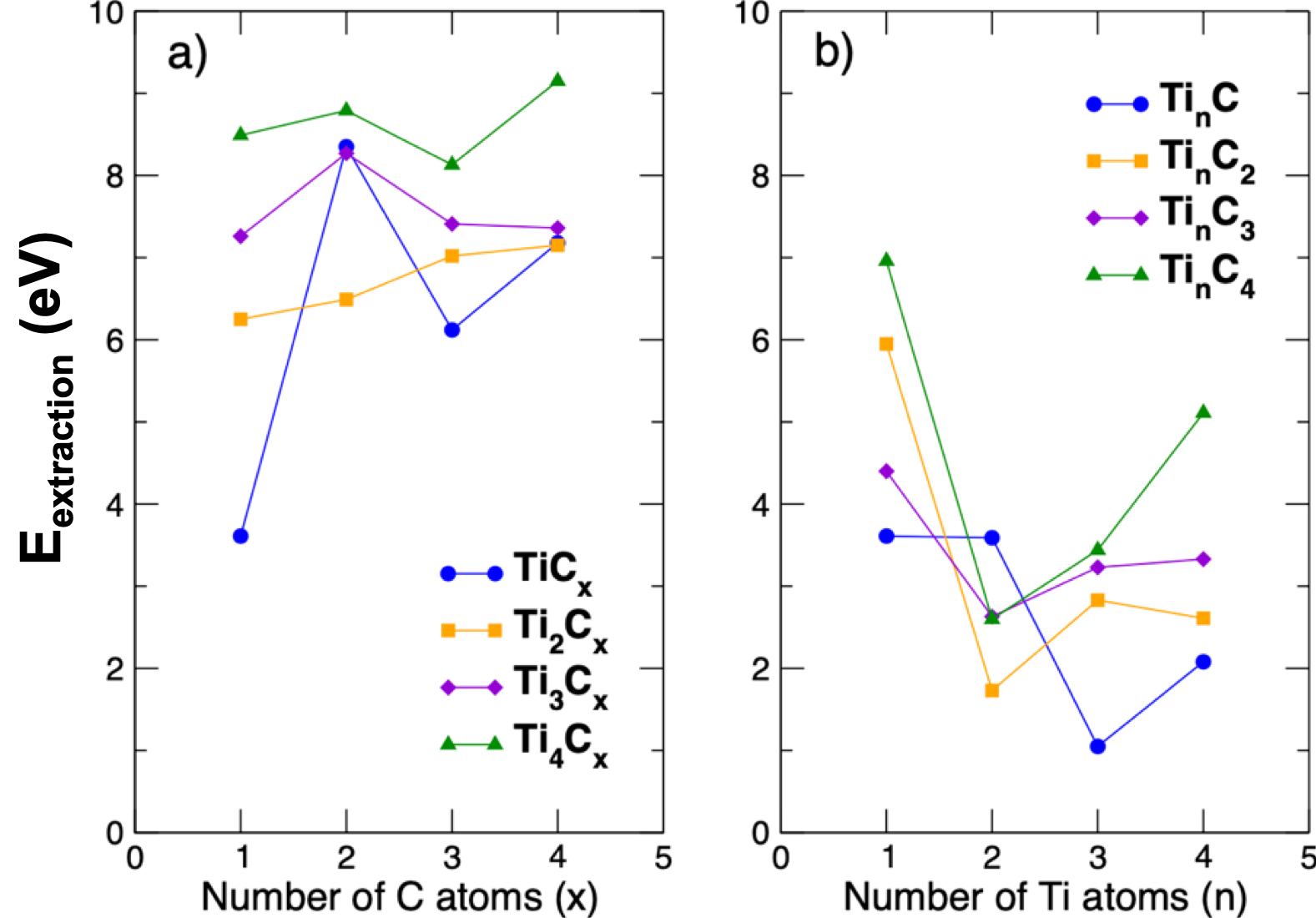}
\caption{(a) Energy required to extract one C atom from Ti$_n$C$_x$ clusters with n = 1--4 and x = 1--4. The horizontal axis indicates the number of C atoms in the initial state of the cluster. Line segments connect clusters with the same number of Ti atoms. (b) Energy required to extract one Ti atom. The horizontal axis indicates the number of Ti atoms in the initial state of the cluster. Line segments connect clusters with the same number of C atoms. }
\label{Fig2}
\end{figure}

Regarding the absorption properties of Ti$_n$C$_x$ clusters, ~\cite{ref18} were the first to measure the infrared absorption spectra of Ti$_8$C$_{12}$ and some other Ti$_n$C$_x$ clusters using the resonance-enhanced multiple photon ionization (REMPI) technique. ~\cite{ref19} calculated the infrared-absorption spectrum of the Ti$_8$C$_{12}$, and we investigated the optical absorption spectrum of this metcar~\citep{ref20,ref21} using time-dependent density functional theory (TD-DFT). Since the ground-state structure of Ti$_8$C$_{12}$ and related metcars had been under debate in the literature, optical absorption spectra of several isomeric structures were studied~\citep{ref21}. All the isomers have a region of high absorption between 10 and 13 eV that was interpreted as a collective electronic excitation. This feature, which had been experimentally noticed by~\cite{ref22}, provides an internal heat bath allowing for the delayed ionization and delayed ion emission observed in these metcars.

Apart from the spectrum of the highly stable Ti$_8$C$_{12}$ cluster, the optical absorption of other Ti-C clusters of relevance in astrochemistry~\citep{ref1} has not been previously addressed. With this in mind, we have used TD-DFT to calculate the visible and near infrared photoabsorption spectrum of small Ti$_n$C$_x$ clusters, with n = 1--4 and x = 1--4, and some selected larger clusters: Ti$_3$C$_8$, Ti$_4$C$_8$, Ti$_6$C$_{13}$, Ti$_7$C$_{13}$, Ti$_8$C$_{12}$, Ti$_9$C$_{15}$, and Ti$_{13}$C$_{22}$. The group of small clusters has been chosen for two reasons: i) trends can be studied as a function of the number of Ti and C atoms in the cluster, and ii) those clusters are involved in the initial stages of Ti-C dust nucleation. On the other hand, the selected larger clusters correspond to clear abundance maxima in the mass spectra obtained in laboratory experiments by~\cite{ref15} and~\cite{ref17}. The calculated optical spectra could serve as basis for the identification of these species in space. Moreover, in addition to the rich optical absorption features, except the highly symmetric clusters, the Ti-C clusters present sizable dipole moments, and therefore could also be identified through radioastronomical observations. We note, however, that for the large clusters the rotational spectra are likely to be crowded by lines, hindering its identification.
 
The structure of the manuscript is as follows. The methodology used in the calculations is presented in Section \ref{section2}. The structure and stability of the clusters are discussed in Section \ref{section3}. Then, Section \ref{section4} presents the absorption spectra and the conclusions are described in Section \ref{section5}. 

\begin{figure*}
\includegraphics[width=\textwidth]{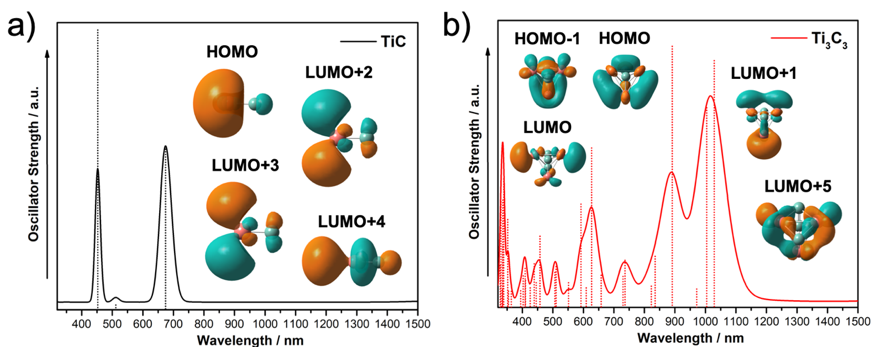}
\caption{Photoabsorption spectrum of TiC and Ti$_3$C$_3$. The continuous curves are obtained by broadening the individual lines with gaussian curves of 0.06 eV width. Insets show the spatial distribution of the HOMO and some representative unoccupied electronic orbitals.}
\label{Fig3}
\end{figure*}

\begin{table}
\centering
\caption{Energies (eV), wavelengths (nm) and oscillator strengths of electronic excitations contributing to the absorption spectrum of TiC. The excitation lines at 511.7 and 675.8 nm are doubly degenerate (*). Excitations with very small oscillator strength are not included.}
	\label{Table1}
	\begin{tabular}{lc|cc}
		\hline
		Energy & & & {\it f} \\
		\hline
		eV & nm & & \\
		\hline
		1.84  & 675.8 & & 0.091 \\
		1.84* & 675.8* & & 0.091 \\
		2.42  & 511.7 & & 0.0027 \\
		2.42* & 511.7* & & 0.0027 \\
		2.74  & 452.5 & & 0.156 \\
		\hline
	\end{tabular}
\end{table}

\begin{figure*}
\includegraphics[width=\textwidth]{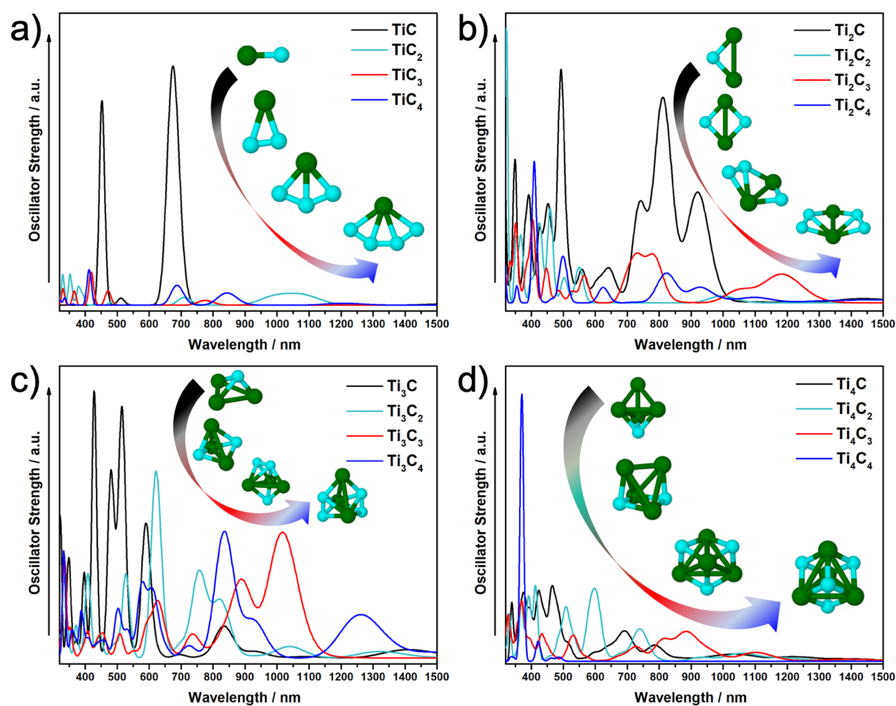}
\caption{Absorption spectra of Ti$_n$C$_x$, n = 1--4 and x = 1--4, in the spectral region from 325 to 1500 nm. The scale of oscillator strenghts is the same in each panel, but it may change from panel to panel.}
\label{Fig4}
\end{figure*}

\begin{figure*}
\includegraphics[width=\textwidth]{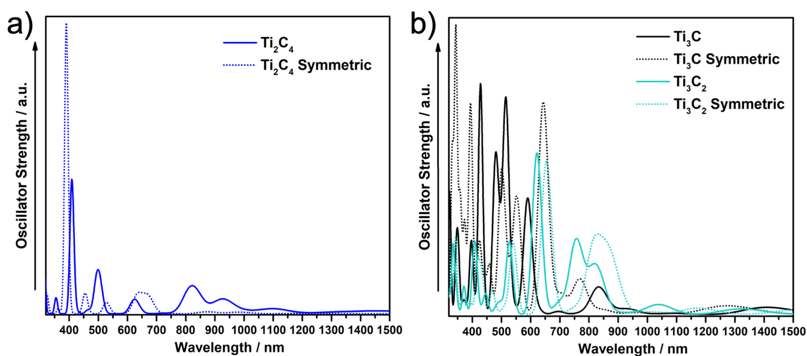}
\caption{Absorption spectra of Ti$_2$C$_4$ in panel (a), and of T$_3$C and Ti$_3$C$_2$ in panel (b). In each case, symmetrized and non-symmetrized versions of the structure are compared.}
\label{Fig5}
\end{figure*}

\section{Computational Approach} \label{section2}

Density functional (DFT)~\citep{ref23,ref24} calculations have been performed to investigate the electronic and atomic structure of the Ti-C clusters. The calculations have been done using the GAUSSIAN16 computer code~\citep{ref25}. For the sake of efficiency and accuracy we have applied two different strategies. For small Ti$_n$C$_x$ clusters with n + x $\leq$ 8, we use an all-electron localized basis formed by linear combinations of Gaussian functions within the cc-pVTZ parametrization~\citep{ref25,ref26}. Exchange and correlation effects are treated with the hybrid functional B3LYP~\citep{ref27,ref28,ref29}, that includes 20\% of exact exchange. Ti$_n$C$_x$ clusters with n + x > 10 are large enough to require the use of pseudo-potentials -- in the present case, the Stuttgart-Dresden Effective Core Potentials (ECPs) implemented in the GAUSSIAN16 code~\citep{ref25} -- for the Ti atoms, keeping the all-electron cc-pVTZ basis for the C atoms. This approach allows a reliable calculation of the lowest energy equilibrium geometries and enthalpies of atomization for all the Ti-C clusters~\citep{ref1}.

For each cluster Ti$_n$C$_x$, a large number of different atomic arrangements were initially prepared, and each of those structures was relaxed and fully optimized. A number of isomeric structures were then discovered (some of the original atomic arrangements relax to the same isomer) and the isomer with the lowest energy was identified as the ground state structure. In the case of the small Ti$_n$C$_x$ clusters (n, x=1--4), all the inequivalent structural configurations (linear, planar, ring-like, three-dimensional polyhedra, etc.) and permutational isomers~\citep{Lopez1996}, that is, the same underlying structure but different distributions of the Ti and C atoms over the available sites, and different spin multiplicities, have been investigated. The calculations delivered results in agreement to those presented in previous papers~\citep{ref39,ref1}. This strategy is not feasible for the clusters with n+x > 10, because of the enormous number of structural and permutational isomers existing. To search for the ground-state structure of these large clusters we started from the geometries already published and other plausible candidate structures. In this case we first relaxed the structures and, as a second step, those structures were shaken and distorted, changing the bond-lengths and angles by amounts up to 15\%. The calculations yielded final structures very similar to those already published. As a step beyond, for some selected large clusters (Ti$_3$C$_8$ and Ti$_7$C$_{13}$) we performed DFT molecular dynamics simulations lasting for up to 1 ps within the microcanonical (NVE) ensemble, at a thermal bath temperature of 300 K, and no phase transitions were observed towards new more stable structures. The methods described allow us to be confident that we have found the true lowest energy structures for the small clusters. In the case of large clusters, although we do not have certitude of having discovered the true lowest energy structures, we are confident of at least having found structures with energies very close to the true ground-state configurations. 
   
The photoabsorption excitations have been calculated using time-dependent density functional theory (TD-DFT)~\citep{ref30,ref31,ref32,ref33,ref34,ref35}, implemented in GAUSSIAN16~\citep{ref25}. A treatment of the first order (linear) response of the system to the external electromagnetic field permits obtaining excitation energies and oscillator strengths from TD-DFT. The dynamic electric polarizability, $\alpha(\omega)$, as a function of the photon frequency can be expressed as
\begin{equation} \label{eq0}
\alpha(\omega)=\sum_{I}\frac{f_I}{\omega_I^{2}-\omega^{2}}.
\end{equation}
That is, the poles of the function $\alpha(\omega)$ provide the excitation energies $\omega_I$  and the residues determine the corresponding oscillator strenghts $f_I$. Using a finite orbital basis set to represent the dynamic polarizability in tensorial form, then the excitation energies and oscillator strenghts can be obtained by solving a matrix eigenvalue equation~\citep{ref35,Jamorski1996,Casida1998}. The orbital basis set is formed from the occupied and unoccupied single-particle wave functions obtained in a previous ordinary (static) DFT calculation with the cluster in its ground state structure (or in an isomeric structure if the absorption spectrum of an isomer is desired). As a result of the TD-DFT calculation, the excitations are obtained as superposition of single-particle-hole transitions between the one-electron occupied and unoccupied states of the basis. That is, the excited state  is a linear combination of single-excited Slater determinants. We have introduced an optimal number of unoccupied states in the basis to accurately obtain excitations with energies up to 5 eV (or, equivalently, above 250 nm), including singlet and triplet excitations.
This requires including between 80 and 120 unoccupied single-particle states in the basis, depending on each specific cluster.

Importantly, in order to achieve an improved description of the excitations within the TD-DFT we have used the long-range exchange-corrected functional CAM-B3LYP~\citep{ref36}, which includes a fraction (19\%) of exact exchange at short range, very similar to the 20\% fraction in the B3LYP, but tends to a limit of 65\% of exact exchange at long range. CAM-B3LYP substantially improves the accuracy in the calculation of the excitations~\citep{ref37}, and appears to provide in some cases a similar accuracy compared to ab initio G$_0$W$_0$ and Bethe-Salpeter equation approaches~\citep{ref38}. It is interesting to remark that the similar short-range implementation of exchange-correlation used in B3LYP and CAM-B3LYP leads to the same ground-state geometries.

The adsorption spectra shown in the Figures of Section~\ref{section4} and in the two Appendices cover a range of wavelengths between 350 and 1500 nm for the small clusters, and between 400 and 1500 nm for the large clusters, although a few excitations corresponding to lower or larger photon energies are also included for some clusters in the Tables of the Appendices. The individual adsorption lines have been broadened by gaussians in order to represent the spectra as continuous curves. The Tables of the Appendices give information about the main single-particle-hole transitions contributing to each excitation in the absorption spectrum. Often, just one particle-hole transition is enough to characterize the excitation, but in other cases two, three or four different particle-hole transitions contribute to the excitation.   

\section{Structure and stability of the clusters} \label{section3}

The lowest energy structures of the Ti$_n$C$_x$ clusters with n = 1--4 and x = 1--4, calculated with the B3LYP functional, are shown in Fig.~\ref{Fig1}. Overall, the structures are in full agreement with those obtained by~\cite{ref39} in a previous DFT calculation. The structural trends can be well understood by focusing on the number n of Ti atoms in the cluster. Clusters with a single Ti atom, TiC$_x$, x = 1--4, are planar and have a fan-like shape~\citep{ref40}. The C atoms form a chain, and all of them are bonded to the Ti atom. The structure of the clusters with n = 2 can be understood by noticing that Ti$_2$C$_2$ is a bent rhombus. The structures of Ti$_2$C$_3$ and Ti$_2$C$_4$ are then obtained by adding one and two C atoms, respectively, in bridge positions between a C and a Ti atom of the rhombus. On the other hand, the structure of Ti$_2$C results by eliminating one C atom from the bent rhombus. The structures of clusters with three Ti atoms can be rationalized by starting with Ti$_3$C$_2$, which is a (deformed) trigonal bipyramid with the Ti atoms forming the equatorial base and the C atoms on the apices. Ti$_3$C$_3$ and Ti$_3$C$_4$ are obtained by capping one and two faces of the bipyramid, respectively, with C atoms; and the structure of Ti$_3$C arises by removing one apex atom from Ti$_3$C$_2$. The structures of clusters with n = 4 derive from that of Ti$_4$C, which is a Ti$_4$ tetrahedron with a face capped by a C atom. Capping two, three and four faces leads to Ti$_4$C$_2$, Ti$_4$C$_3$ and Ti$_4$C$_4$, respectively. Ti$_4$C$_4$ is a cube in which the three nearest neighbors of a C atom are Ti atoms, and vice versa. It can be noticed that the calculated lowest energy structure of pure Ti$_4$ is also a tetrahedron~\citep{ref41} or a distorted tetrahedron~\citep{ref42}. Interestingly, Ti$_4$C$_4$ is a small fragment of the solid TiC compound, which has an fcc-like NaCl structure~\citep{ref43}.

A complementary interpretation of the structures is obtained by noticing that the Ti atoms form a core fragment, and the C atoms tend to surround that fragment. This is incipient in TiC$_x$, where the C atoms start surrounding the Ti atom. The Ti$_2$C$_x$ family begins with a C atom on one side of the Ti dimer. The second C atom goes to the opposite side. Ti$_2$C$_3$ and Ti$_2$C$_4$ just follow the trend, putting a second C atom on each side, in that way forming one and two carbon dimers, respectively. The Ti$_3$C$_x$ family begins with a triangular Ti$_3$ core capped by a C atom. The second C atom caps the other side of the triangle, forming a trigonal bipyramid. Ti$_3$C$_3$ and Ti$_3$C$_4$ follow the trend, with a second C atom on each side of the Ti$_3$ triangle, again forming one and two carbon dimers, respectively. Finally, in the Ti$_4$C$_x$ group, the four faces of a Ti$_4$ tetrahedron are successively capped by C atoms. 
The structures of some selected larger clusters, Ti$_3$C$_8$, Ti$_4$C$_8$, Ti$_6$C$_{13}$, Ti$7$C$_{13}$, Ti$_8$C$_{12}$, Ti$_9$C$_{15}$, and Ti$_{13}$C$_{22}$, are also shown in Fig.~\ref{Fig1}. Ti$_3$C$_8$ displays the intact structure of Ti$_3$C$_2$, surrounded by three carbon dimers attached in bridge-like positions to the Ti--Ti bonds. The orientation of those carbon dimers is near perpendicular to the Ti--Ti bonds. The structure of Ti$_4$C$_8$ is based on that of Ti$_4$C$_4$, but each one of the four C atoms of the external tetrahedron of Ti$_4$C$_4$ has been replaced by a carbon dimer. The presence of carbon dimers is also observed in larger clusters: Ti$_6$C$_{13}$, Ti$_7$C$_{13}$, Ti$_8$C$_{12}$, Ti$_9$C$_{15}$, and Ti$_{13}$C$_{22}$. In fact, carbon dimers have also been noticed above in small clusters: Ti$_2$C$_3$, Ti$_2$C$_4$, Ti$_3$C$_3$ and Ti$_3$C$_4$. The clusters Ti$_6$C$_{13}$ and Ti$_7$C$_{13}$ have the form of empty cages with some carbon dimers capping the cage. Ti$_8$C$_{12}$, known as metallocarbohedrene, is also a cage~\citep{ref11,ref12,ref13,ref14}. Due to its high stability, this cluster has been exhaustively investigated~\citep{ref14,ref21}, and a number of isomers exist close in energy. The structure we obtained has T$_d$ symmetry, also found in earlier works~\citep{ref14,ref21,ref44,ref45}, although the structure of the metcar is still under discussion~\citep{Megha2021}. Finally, Ti$_{13}$C$_{22}$ has a Ti atom in the cluster center.

In general, the structures of Figure~\ref{Fig1} show a Ti cluster (compact in the small titanium carbide clusters and non-compact in the others) surrounded by C atoms, forming C--C dimers in many cases, and only in a few cases larger carbon chains. So, segregation of the C atoms is observed, but the clusters are not large enough to display a clear core-shell structure. Starting with Ti$_4$C$_8$ the clusters form cages, and only in the largest cluster studied, Ti$_{13}$C$_{22}$, an atom appears enclosed at the center of the cage. However, the fcc-like NaCl structure of the macroscopic Ti carbide is not yet achieved for the sizes studied. As an exception, the only cluster that can be viewed as a NaCl-like fragment is Ti$_4$C$_4$. Small silicon carbide clusters, also observed in the dust of AGB stars, have been theoretically investigated by~\cite{Gobrecht2017} using DFT. Although the silicon carbide stoichiometry studied was equiatomic, that is, Si$_n$C$_n$, n = 1-- 16, and the most stable structures are sensitive, for some sizes, to the exchange-correlation functional used in the calculations, it is worth to make a comparison between the trends found in the titanium carbide and silicon carbide clusters. The estructural trends are different, and this is not surprising because C and Si are chemical elements in the same column (electronic configuration s$^{2}$p$^{2}$) of the periodic table, while Ti is a transition metal element (configuration d$^{2}$s$^{2}$). The most stable structures of silicon carbide clusters reveal a marked tendency to form C--C bonds, which produces carbon-segregated substructures; while bulk-like isomers with alternating Si--C bonds are much less stable. Complete and incomplete pentagons and hexagons are found in those carbon substructures, and the Si atoms attach to the periphery of those carbon substructures as individual Si atoms or small Si clusters. For n > 10, those segregated structures adopt the form of empty cages (the cage is not closed for Si$_{10}$C$_{10}$). So, formation of cages starts earlier in the titanium carbide clusters as compared to silicon carbide. Si$_{12}$C$_{12}$ and Si$_{16}$C$_{16}$ are, however, special because the cages are based on an arrangement of alternating Si--C bonds, instead of segregated substructures. Something common to silicon carbide and titanium carbide clusters is that bulk-like structures are not present in the small or medium size (up to three dozen atoms) clusters.

The stability of the clusters can be explored by considering the energies of the dissociation reactions:
\begin{eqnarray} \label{eqs12}
Ti_nC_x &\rightarrow& Ti_nC_{x-1} + C, \\
Ti_nC_x &\rightarrow& Ti_{n-1}C_x + Ti,
\end{eqnarray}
that is, the energies required to remove a C or a Ti atom from the cluster. Those dissociation energies are plotted in Fig.~\ref{Fig2} for the clusters with n = 1--4 and x = 1--4. Panel (a) collects the energies to remove a C atom, and panel (b) the energies to remove a Ti atom. The first observation is that removing C atoms requires, in general, more energy than removing Ti atoms. The energy to release C atoms increases with the number of Ti atoms in the cluster. That is, the order of the curves is TiC$_x$, Ti$_2$C$_x$, Ti$_3$C$_x$, Ti$_4$C$_x$. The TiC$_2$ cluster represents a clear exception. The energy to remove one C atom from this cluster is substantially higher than the corresponding extraction energies in the other clusters of the n=1 family, namely TiC, TiC$_3$, TiC$_4$, and indicates that a C$_2$ dimer attached to a Ti atom is a very stable unit. In fact, as pointed out above, this unit can be identified in the structure of many clusters in Fig.~\ref{Fig1}. In the other families, n = 2, 3, 4, the C extraction energy shows a moderate variation with the initial number x of C atoms. Concerning the extraction of Ti, the only clear trend is that in each x family the highest extraction energy occurs for TiC$_x$. Most clusters require a large amount of energy to extract a C atom, and a substantially lower energy to extract a Ti atom. But only a few have both large Ti and C extraction energies. These are TiC$_2$, TiC$_4$ and Ti$_4$C$_4$, which are the most stable clusters, and to a less extent, also TiC$_3$. Interestingly, TiC$_2$ is predicted as the most abundant Ti-C molecular species at distances shorter than 2 stellar radii in C-rich AGB stars~\citep{ref1}, which is very likely related to its stability.

\section{Absorption spectra} \label{section4}

The absorption spectra have been calculated by TD-DFT using the CAM-B3LYP functional. In general, it is not possible to assign each excitation to an electronic transition between a specific occupied single-particle state and a specific unoccupied single-particle state. Instead, each excitation can be viewed as a combination of single-particle transitions between the occupied and unoccupied states forming the basis used for the TD-DFT calculations, that is, the single particle states from the static time-independent DFT calculation. 

The spectrum of TiC in the 325 nm to 1500 nm spectral region is shown in panel (a) of Fig.~\ref{Fig3}. The position and height of the vertical lines indicate the absorption wavelengths and the oscillator strengths, respectively. The energy, wavelength and oscillator strength of the most prominent excitations are given in Table~\ref{Table1}. The spectrum of TiC is dominated by two absorption lines in the visible region, at 452.5 and 675.8 nm, respectively. There is also a doubly degenerate absorption feature at 511.7 nm, although with very low oscillator strength. The line at 675.8 nm is degenerate, comprised of two modes with the same energy and oscillator strength. Broadening the spectral lines with Gaussian curves of 0.06 eV width produces the spectrum given by the continuous curve in Fig.~\ref{Fig3}a. The insets show the spatial distribution of the HOMO (highest occupied molecular orbital) along with some unoccupied electronic orbitals that are the main contributing ones to the electronic excitations of TiC. The HOMO, LUMO+2, and LUMO+3 orbitals are predominantly localized on the Ti atom, whereas LUMO+4 is distributed over the two atoms (the notation LUMO, LUMO+1, ...  stands for the lowest unoccupied molecular orbital, the next unoccupied molecular orbital, etc). For comparison, the absorption spectrum of Ti$_3$C$_3$ is shown in panel (b) of Fig.~\ref{Fig3}. In this case, the spectrum is more complex, exhibiting numerous absorption lines. The excitations with highest oscillator strength occur in the near-infrared, with several groups of closely spaced excitations appearing also in the visible spectral region. 

The broadened absorption spectra of Ti$_n$C$_x$, n = 1--4 and x = 1--4, for wavelengths from 325 to 1500 nm, are shown in Fig.~\ref{Fig4}. Each individual panel stands for a specific value of the number n of Ti atoms. The individual spectra, including the vertical lines representing the oscillator strength of the excitations are collected in Appendix~\ref{apA}. All the clusters present several excitations in the visible region, and the form of the broadened spectrum is motivated by the excitations with the highest oscillator strength, as one can see in the two examples shown in Fig.~\ref{Fig3}, TiC and Ti$_3$C$_3$. In panels (a), (b) and (c) of Fig.~\ref{Fig4}, the highest oscillator strengths occur for the cluster with a single C atom. However, the differences in oscillator strength between the clusters with one and more C atoms decreases from panel (a) to panel (b), and then to panel (c); that is, as the number of Ti atoms in the cluster increases. Panel (d) shows a different behavior, because Ti$_4$C$_4$ is a particular case. This cluster has a symmetric cube structure, and the excitation spectrum is concentrated on two lines, at 370.1 and 421.7 nm (3.35 and 2.94 eV). Each one of those lines is triply degenerate (three different excitations with the same energy), but the oscillator strength of the excitations at 370.1 nm is one order of magnitude larger than those at 421.7 nm.  

\begin{figure}
\includegraphics[width=\columnwidth]{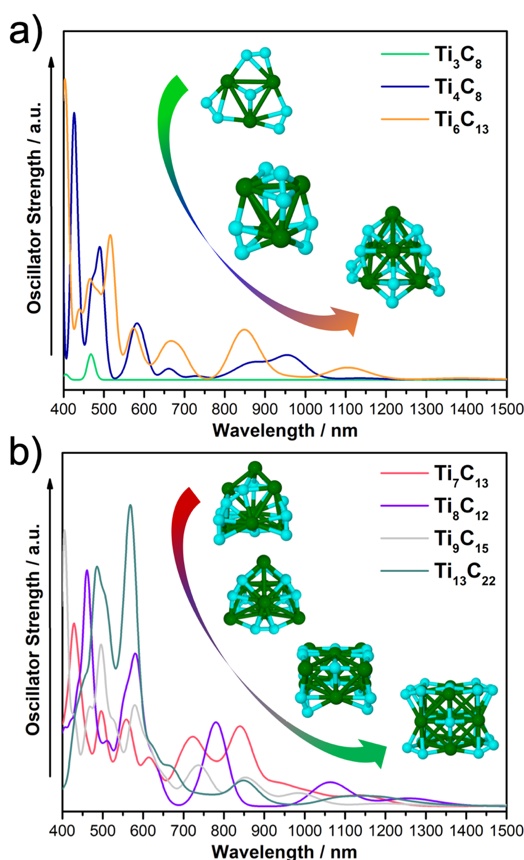}
\caption{Absorption spectra of Ti$_3$C$_8$, Ti$_4$C$_8$, Ti$_6$C$_{13}$, Ti$7$C$_{13}$, Ti$_8$C$_{12}$, Ti$_9$C$_{15}$, and Ti$_{13}$C$_{22}$.}
\label{Fig6}
\end{figure}

The photoabsorption spectrum is sensitive to the cluster geometry~\citep{ref46}, and this effect is analyzed in Fig.~\ref{Fig5}. Absorption spectra are presented for Ti$_2$C$_4$ in panel (a); and for Ti$_3$C and Ti$_3$C$_2$ in panel (b). Since, in general, the structure of the clusters shows distortions with respect to a perfect symmetric structure, in Fig.~\ref{Fig5} a comparison is provided between the spectra calculated for the lowest energy structures of Fig.~\ref{Fig1} and for the corresponding ideal symmetric structures. Ti$_2$C$_4$ is a bent Ti$_2$C$_2$ rhombus with two sides capped by C atoms. Ti$_3$C$_2$ is a trigonal bipyramid, and Ti$_3$C is obtained by eliminating one apex atom from the bipyramid. Those structures are clearly recognized in the lowest energy structures, because the distortions from the ideal structures are small. The smallest distortion occurs for Ti$_3$C$_2$, the difference in energy between the lowest energy structure and the symmetrized version  being only 0.035 eV, and indeed the two absorption spectra are quite close. On the other hand, the distortions are larger in the other two clusters. The energy differences between the lowest energy structures and the symmetrized versions are 0.254 and 0.370 eV for Ti$_3$C and Ti$_2$C$_4$, respectively, and the corresponding absorption spectra show appreciable differences in the position and height of the absorption peaks.

The absorption spectra of the large clusters, Ti$_3$C$_8$, Ti$_4$C$_8$, Ti$_6$C$_{13}$, Ti$7$C$_{13}$, Ti$_8$C$_{12}$, Ti$_9$C$_{15}$, and Ti$_{13}$C$_{22}$, are collected in Fig.~\ref{Fig6}. The excitations with higher oscillator strengths are concentrated in the visible region, but absorptions in the near IR region are also appreciable. The individual spectra, including the vertical lines representing the oscillator strength of the excitations are collected in Appendix~\ref{apB}. Both Ti$_7$C$_{13}$ and Ti$_8$C$_{12}$ have twenty atoms, but the spectra of these two clusters are quite distinct due to the different relative composition and the structural differences between the two clusters. Ti$_8$C$_{12}$ shows a Ti$_8$ structure decorated by well separated carbon dimers. In contrast, chains of several C atoms are found in Ti$_7$C$_{13}$, and the Ti fragment in Ti$_7$C$_{13}$ is less compact than in Ti$_8$C$_{12}$.
     
At this point it is important to remark that a comparison of our computed absorption profiles with experimental absorption spectra of titanium carbide clusters is a difficult task because, to the best of our knowledge, experiments have only been performed for large nanoparticles, 40--50 nanometers in size~\citep{Kimura2003} or much larger~\citep{Henning2001}, and for solid carbides~\citep{Pfluger1984,Koide1990}. The experiments on nanoparticles by~\cite{Kimura2003} and~\cite{Henning2001} have focused on the infrared spectra, related to vibrations. On the other hand, electron-energy-loss spectroscopy~\citep{Pfluger1984} and reflectance~\citep{Koide1990} experiments on solid carbides sampled a broad range of energies, up to 40 eV. As can be appreciated in Figures~\ref{Fig4} and~\ref{Fig6}, the absorption spectrum in the visible region is highly sensitive to cluster size and relative concentration of Ti and C. But a qualitative comparison with the TiC bulk can be established. The measured reflectance in the range of energies between 0 and 40 eV shows an overall non-smooth decrease (that is, an increase in the absorption) as the photon energy increases~\citep{Pfluger1984,Koide1990}. This correlates with the behavior of the calculated absorption in Figures~\ref{Fig4} and~\ref{Fig6}, small or negligible in the region of long wavelengths, and increasing substantially as the wavelength decreases.
 
The stoichiometric TiC carbide bulk and large nanoparticles crystallize in the rocksalt (or NaCl) structure. It was pointed out above that the Ti$_4$C$_4$ cluster is a small fragment of the rocksalt lattice, so one can be tempted to establish some relation between the calculated spectrum of Ti$_4$C$_4$ and the measured spectra of large TiC nanoparticles and the bulk carbide. The visible spectrum of Ti$_4$C$_4$ shows features at 421.7 nm (2.94 eV) and 370.1 nm (3.35 eV) in Figures~\ref{Fig4} and~\ref{FigA16}. ~\cite{Koide1990} compared the reflectance spectrum of near stoichiometric TiC in the low energy region (0--5 eV) to previous reflectance measurements by other workers, concluding that all of the reported data agree only on the existence of a reflectance peak at approximately 3.2 eV. That feature correlates reasonably well with the region of no absorption, centered at 3.15 eV and flanked by absorption peaks at 2.94 eV and 3.35 eV, in the spectrum of Ti$_4$C$_4$. Even if suggestive, this apparent correlation should not be overstated because of the difference between bulk TiC and the Ti$_4$C$_4$ cluster.

Apart from the optical absorption spectra of the clusters, we have calculated the magnitudes of the electric dipole moments of all the clusters, which are summarized in Tables~\ref{Table2} and~\ref{Table3}. Not surprisingly, fully symmetric clusters present null electric dipole moments, and the dipole moment increases as the cluster structure is less symmetric. We would like to note the large dipole moments for the clusters with a single Ti atom (TiC$_x$), which in principle would allow the identification of these species by radioastronomical observations. Chemical equilibrium calculations predict the presence of TiC$_x$ clusters with x = 1--4 in the innermost regions of C-rich AGBs, with TiC$_2$ being the most abundant, although all of them present very low abundances~\citep{ref1}. Nevertheless, there are evidences of nonthermal equilibrium in the inner wind of AGB stars~\citep{ref5,ref47,ref48} implying that the abundances of TiC clusters in the inner envelopes might differ notably from those calculated under thermal equilibrium conditions. Regarding the large Ti$_n$C$_x$ clusters, Ti$_8$C$_{12}$ and Ti$_{13}$C$_{22}$ present permanent dipole moments. The former has been suggested as precursor of TiC dust and the calculated abundances of both clusters at 3-10 stellar radii~\citep{ref1} make them ideal candidates for its detection by radioastronomical observations.

\begin{table}
	\centering
	\caption{Electric dipole moments of small clusters (debye).}
	\label{Table2}
	\begin{tabular}{lclrlclc} 
		\hline
		{\bf TiC}         & 6.66 & {\bf Ti$_2$C}         & 2.76 & {\bf Ti$_3$C}         & 2.40 & {\bf Ti$_4$C}         & 2.03 \\
		{\bf TiC$_2$} & 7.18 & {\bf Ti$_2$C$_2$} & 0.00 & {\bf Ti$_3$C$_2$} & 0.86 & {\bf Ti$_4$C$_2$} & 0.07 \\
		{\bf TiC$_3$} & 6.47 & {\bf Ti$_2$C$_3$} & 5.11 & {\bf Ti$_3$C$_3$} & 1.96 & {\bf Ti$_4$C$_3$} & 1.97 \\
		{\bf TiC$_4$} & 4.93 & {\bf Ti$_2$C$_4$} & 5.72 & {\bf Ti$_3$C$_4$} & 1.17 & {\bf Ti$_4$C$_4$} & 0.00 \\
		\hline
	\end{tabular}
\end{table}

\begin{table}
	\centering
	\caption{Electric dipole moments of large clusters (debye).}
	\label{Table3}
	\begin{tabular}{ccccccc} 
		\hline
		{\bf Ti$_3$C$_8$} & {\bf Ti$_4$C$_8$} & {\bf Ti$_6$C$_{13}$} & {\bf Ti$_7$C$_{13}$} & {\bf Ti$_8$C$_{12}$} & {\bf Ti$_9$C$_{15}$} & {\bf Ti$_{13}$C$_{22}$} \\
		0.00 & 2.89 & 9.90 & 6.84 & 0.75 & 14.44 & 2.17 \\
		\hline
	\end{tabular}
\end{table}

\section{Conclusions} \label{section5}

Small clusters formed by Ti and C atoms are of high interest in astrochemistry, since TiC is considered to be the first condensate in the CSEs of C-rich AGBs providing the nucleation seeds for cosmic dust growth. However, the condensation of TiC leading to the formation of TiC dust grains is far from being understood. In particular, the molecular precursors are still unknown and the identification of Ti$_n$C$_x$ clusters in space could provide valuable information on the mechanism of dust growth in evolved stars.  

Here, using the density functional formalism, with the B3LYP approximation for electronic exchange and correlation effects, we have investigated the structure and stability of small Ti$_n$C$_x$ clusters, with n = 1--4 and x = 1--4, and a few selected larger clusters: Ti$_3$C$_8$, Ti$_4$C$_8$, Ti$_6$C$_{13}$, Ti$7$C$_{13}$, Ti$_8$C$_{12}$, Ti$_9$C$_{15}$, and Ti$_{13}$C$_{22}$. Using the time-dependent density functional theory, we have also calculated the absorption spectra of Ti$_n$C$_x$ clusters, which show excitation peaks in the visible and near IR spectral regions. In addition, most of the clusters investigated have sizable electric dipole moments allowing its observation by rotational spectroscopy. Our results may help in the detection of TiC cluster species in AGB and post-AGB stars.

\section*{Acknowledgements}

S.G.-V. thanks the MINECO for a FPU predoctoral fellowship (FPU17/04908). The work of J.A.A. was supported by Ministerio de Ciencia e Innovaci\'on of Spain (grant PID2019-104924RB-I00), Junta de Castilla y Le\'on (grant VA021G18), and Universidad de Valladolid (GIR Nanostructure Physics). G.S. and J.I.M. acknowledge financial support from Spanish MINECO (grant MAT2017-85089-C2-1-R), European Research Council (ERC) under contract ERC-2013-SYG-610256 NANOCOSMOS, and Comunidad de Madrid via ``Programa de Investigaci\'on Tecnolog\'{\i}as'' 2018 (FOTOART-CM S2018/NMT-4367).

\section*{Data Availability}

The data underlying this article will be shared on reasonable request to the corresponding author.



\bibliographystyle{mnras}
\bibliography{biblio.bib} 

\begin{thebibliography}{}
\makeatletter
\relax
\def\mn@urlcharsother{\let\do\@makeother \do\$\do\&\do\#\do\^\do\_\do\%\do\~}
\def\mn@doi{\begingroup\mn@urlcharsother \@ifnextchar [ {\mn@doi@}
  {\mn@doi@[]}}
\def\mn@doi@[#1]#2{\def\@tempa{#1}\ifx\@tempa\@empty \href
  {http://dx.doi.org/#2} {doi:#2}\else \href {http://dx.doi.org/#2} {#1}\fi
  \endgroup}
\def\mn@eprint#1#2{\mn@eprint@#1:#2::\@nil}
\def\mn@eprint@arXiv#1{\href {http://arxiv.org/abs/#1} {{\tt arXiv:#1}}}
\def\mn@eprint@dblp#1{\href {http://dblp.uni-trier.de/rec/bibtex/#1.xml}
  {dblp:#1}}
\def\mn@eprint@#1:#2:#3:#4\@nil{\def\@tempa {#1}\def\@tempb {#2}\def\@tempc
  {#3}\ifx \@tempc \@empty \let \@tempc \@tempb \let \@tempb \@tempa \fi \ifx
  \@tempb \@empty \def\@tempb {arXiv}\fi \@ifundefined
  {mn@eprint@\@tempb}{\@tempb:\@tempc}{\expandafter \expandafter \csname
  mn@eprint@\@tempb\endcsname \expandafter{\@tempc}}}

\bibitem[\protect\citeauthoryear{Adamo \& Jacquemin}{Adamo \&
  Jacquemin}{2013}]{ref33}
Adamo C.,  Jacquemin D.,  2013, \mn@doi [Chem. Soc. Rev.] {10.1039/C2CS35394F},
  42, 845

\bibitem[\protect\citeauthoryear{{Ag\'undez, M.}, {Mart\'{\i}nez, J. I.}, {de
  Andres, P. L.}, {Cernicharo, J.}  \& {Mart\'{\i}n-Gago, J. A.}}{{Ag\'undez,
  M.} et~al.}{2020}]{ref1}
{Ag\'undez, M.} {Mart\'{\i}nez, J. I.} {de Andres, P. L.} {Cernicharo, J.}
  {Mart\'{\i}n-Gago, J. A.} 2020, \mn@doi [A\&A] {10.1051/0004-6361/202037496},
  637, A59

\bibitem[\protect\citeauthoryear{Becke}{Becke}{1993a}]{ref27}
Becke A.~D.,  1993a, \mn@doi [The Journal of Chemical Physics]
  {10.1063/1.464304}, 98, 1372

\bibitem[\protect\citeauthoryear{Becke}{Becke}{1993b}]{ref28}
Becke A.~D.,  1993b, \mn@doi [The Journal of Chemical Physics]
  {10.1063/1.464913}, 98, 5648

\bibitem[\protect\citeauthoryear{{Bernatowicz}, {Amari}, {Zinner}  \&
  {Lewis}}{{Bernatowicz} et~al.}{1991}]{ref2}
{Bernatowicz} T.~J.,  {Amari} S.,  {Zinner} E.~K.,   {Lewis} R.~S.,  1991,
  \mn@doi [\apjl] {10.1086/186054}, \href
  {https://ui.adsabs.harvard.edu/abs/1991ApJ...373L..73B} {373, L73}

\bibitem[\protect\citeauthoryear{Bernatowicz, Cowsik, Gibbons, Lodders, Fegley,
  Amari  \& Lewis}{Bernatowicz et~al.}{1996}]{ref3}
Bernatowicz T.~J.,  Cowsik R.,  Gibbons P.~C.,  Lodders K.,  Fegley B.,  Amari
  S.,   Lewis R.~S.,  1996, \mn@doi [The Astrophysical Journal]
  {10.1086/178105}, 472, 760

\bibitem[\protect\citeauthoryear{Bourass et~al.,}{Bourass et~al.}{2015}]{ref37}
Bourass M.,  et~al., 2015, {Journal of Materials and Environmental Science}, 6,
  1542

\bibitem[\protect\citeauthoryear{Casida}{Casida}{1995}]{ref35}
Casida M.~E.,  1995, Time-Dependent Density Functional Response Theory for
  Molecules.
World Scientific, pp 155--192

\bibitem[\protect\citeauthoryear{Casida, Jamorski, Casida  \& Salahub}{Casida
  et~al.}{1998}]{Casida1998}
Casida M.~E.,  Jamorski C.,  Casida K.~C.,   Salahub D.~R.,  1998, \mn@doi [The
  Journal of Chemical Physics] {10.1063/1.475855}, 108, 4439

\bibitem[\protect\citeauthoryear{Castro, Marques, Alonso, Bertsch, Yabana  \&
  Rubio}{Castro et~al.}{2002}]{ref46}
Castro A.,  Marques M. A.~L.,  Alonso J.~A.,  Bertsch G.~F.,  Yabana K.,
  Rubio A.,  2002, \mn@doi [The Journal of Chemical Physics]
  {10.1063/1.1430737}, 116, 1930

\bibitem[\protect\citeauthoryear{Castro, Marques, Alonso  \& Rubio}{Castro
  et~al.}{2004}]{ref32}
Castro A.,  Marques M. A.~L.,  Alonso J.~A.,   Rubio A.,  2004, Journal of
  Computational and Theoretical Nanoscience, 1, 231

\bibitem[\protect\citeauthoryear{Chang \& Graham}{Chang \&
  Graham}{1966}]{ref43}
Chang R.,  Graham L.~J.,  1966, \mn@doi [Journal of Applied Physics]
  {10.1063/1.1707923}, 37, 3778

\bibitem[\protect\citeauthoryear{{Cherchneff, I.}}{{Cherchneff,
  I.}}{2006}]{ref47}
{Cherchneff, I.} 2006, \mn@doi [A\&A] {10.1051/0004-6361:20064827}, 456, 1001

\bibitem[\protect\citeauthoryear{{Cherchneff, I.}}{{Cherchneff,
  I.}}{2012}]{ref48}
{Cherchneff, I.} 2012, \mn@doi [A\&A] {10.1051/0004-6361/201118542}, 545, A12

\bibitem[\protect\citeauthoryear{Chigai, Yamamoto  \& Kozasa}{Chigai
  et~al.}{1999}]{ref4}
Chigai T.,  Yamamoto T.,   Kozasa T.,  1999, \mn@doi [The Astrophysical
  Journal] {10.1086/306628}, 510, 999

\bibitem[\protect\citeauthoryear{Chigai, Yamamoto, Kaito  \& Kimura}{Chigai
  et~al.}{2003}]{ref7}
Chigai T.,  Yamamoto T.,  Kaito C.,   Kimura Y.,  2003, \mn@doi [The
  Astrophysical Journal] {10.1086/368299}, 587, 771

\bibitem[\protect\citeauthoryear{Dance}{Dance}{1992}]{ref44}
Dance I.,  1992, \mn@doi [J. Chem. Soc.{,} Chem. Commun.]
  {10.1039/C39920001779}, pp 1779--1780

\bibitem[\protect\citeauthoryear{Dance}{Dance}{1996}]{ref45}
Dance I.,  1996, \mn@doi [Journal of the American Chemical Society]
  {10.1021/ja9608196}, 118, 6309

\bibitem[\protect\citeauthoryear{Dunning}{Dunning}{1989}]{ref26}
Dunning T.~H.,  1989, \mn@doi [The Journal of Chemical Physics]
  {10.1063/1.456153}, 90, 1007

\bibitem[\protect\citeauthoryear{Frisch et~al.,}{Frisch et~al.}{2009}]{ref25}
Frisch M.~J.,  et~al., 2009, Gaussian09 {R}evision {E}.01

\bibitem[\protect\citeauthoryear{Gail \& Sedlmayr}{Gail \&
  Sedlmayr}{2013}]{ref5}
Gail H.-P.,  Sedlmayr E.,  2013, Physics and Chemistry of Circumstellar Dust
  Shells.
Cambridge Astrophysics, Cambridge University Press,
  \mn@doi{10.1017/CBO9780511985607}

\bibitem[\protect\citeauthoryear{Gobrecht, Cristallo, Piersanti  \&
  Bromley}{Gobrecht et~al.}{2017}]{Gobrecht2017}
Gobrecht D.,  Cristallo S.,  Piersanti L.,   Bromley S.~T.,  2017, \mn@doi [The
  Astrophysical Journal] {10.3847/1538-4357/aa6db0}, 840, 117

\bibitem[\protect\citeauthoryear{Gueorguiev \& Pacheco}{Gueorguiev \&
  Pacheco}{2002}]{ref19}
Gueorguiev G.~K.,  Pacheco J.~M.,  2002, \mn@doi [Phys. Rev. Lett.]
  {10.1103/PhysRevLett.88.115504}, 88, 115504

\bibitem[\protect\citeauthoryear{Guo, Kerns  \& Castleman}{Guo
  et~al.}{1992a}]{ref11}
Guo B.~C.,  Kerns K.~P.,   Castleman A.~W.,  1992a, \mn@doi [Science]
  {10.1126/science.255.5050.1411}, 255, 1411

\bibitem[\protect\citeauthoryear{Guo, Wei, Purnell, Buzza  \& Castleman}{Guo
  et~al.}{1992b}]{ref13}
Guo B.~C.,  Wei S.,  Purnell J.,  Buzza S.,   Castleman A.~W.,  1992b, \mn@doi
  [Science] {10.1126/science.256.5056.515}, 256, 515

\bibitem[\protect\citeauthoryear{Helden, Tielens, Heijnsbergen, Duncan, Hony,
  Waters  \& Meijer}{Helden et~al.}{2000}]{ref6}
Helden G.~v.,  Tielens A. G. G.~M.,  Heijnsbergen D.~v.,  Duncan M.~A.,  Hony
  S.,  Waters L. B. F.~M.,   Meijer G.,  2000, \mn@doi [Science]
  {10.1126/science.288.5464.313}, 288, 313

\bibitem[\protect\citeauthoryear{Henning \& Mutschke}{Henning \&
  Mutschke}{2001}]{Henning2001}
Henning T.,  Mutschke H.,  2001, \mn@doi [Spectrochimica Acta Part A: Molecular
  and Biomolecular Spectroscopy]
  {https://doi.org/10.1016/S1386-1425(00)00446-7}, 57, 815

\bibitem[\protect\citeauthoryear{Jamorski, Casida  \& Salahub}{Jamorski
  et~al.}{1996}]{Jamorski1996}
Jamorski C.,  Casida M.~E.,   Salahub D.~R.,  1996, \mn@doi [The Journal of
  Chemical Physics] {10.1063/1.471140}, 104, 5134

\bibitem[\protect\citeauthoryear{{Kimura} \& {Kaito}}{{Kimura} \&
  {Kaito}}{2003}]{Kimura2003}
{Kimura} Y.,  {Kaito} C.,  2003, \mn@doi [\mnras]
  {10.1046/j.1365-8711.2003.06675.x}, \href
  {https://ui.adsabs.harvard.edu/abs/2003MNRAS.343..385K} {343, 385}

\bibitem[\protect\citeauthoryear{Kohn \& Sham}{Kohn \& Sham}{1965}]{ref23}
Kohn W.,  Sham L.~J.,  1965, \mn@doi [Phys. Rev.] {10.1103/PhysRev.140.A1133},
  140, A1133

\bibitem[\protect\citeauthoryear{Koide, Shidara, Fukutani, Fujimori, Miyahara,
  Kato, Otani  \& Ishizawa}{Koide et~al.}{1990}]{Koide1990}
Koide T.,  Shidara T.,  Fukutani H.,  Fujimori A.,  Miyahara T.,  Kato H.,
  Otani S.,   Ishizawa Y.,  1990, \mn@doi [Phys. Rev. B]
  {10.1103/PhysRevB.42.4979}, 42, 4979

\bibitem[\protect\citeauthoryear{Largo, Cimas, Redondo, Ray{\'o}n  \&
  Barrientos}{Largo et~al.}{2006}]{ref40}
Largo L.,  Cimas A.,  Redondo P.,  Ray{\'o}n V.~M.,   Barrientos C.,  2006,
  \mn@doi [Chemical Physics] {https://doi.org/10.1016/j.chemphys.2006.09.023},
  330, 431

\bibitem[\protect\citeauthoryear{Laurent, Adamo  \& Jacquemin}{Laurent
  et~al.}{2014}]{ref34}
Laurent A.~D.,  Adamo C.,   Jacquemin D.,  2014, \mn@doi [Phys. Chem. Chem.
  Phys.] {10.1039/C3CP55336A}, 16, 14334

\bibitem[\protect\citeauthoryear{Lee, Yang  \& Parr}{Lee et~al.}{1988}]{ref29}
Lee C.,  Yang W.,   Parr R.~G.,  1988, \mn@doi [Phys. Rev. B]
  {10.1103/PhysRevB.37.785}, 37, 785

\bibitem[\protect\citeauthoryear{L{\'o}pez, Marcos  \& Alonso}{L{\'o}pez
  et~al.}{1996}]{Lopez1996}
L{\'o}pez M.~J.,  Marcos P.~A.,   Alonso J.~A.,  1996, \mn@doi [The Journal of
  Chemical Physics] {10.1063/1.470831}, 104, 1056

\bibitem[\protect\citeauthoryear{Marques \& Gross}{Marques \&
  Gross}{2004}]{ref31}
Marques M. A.~L.,  Gross E. K.~U.,  2004, \mn@doi [Annual Review of Physical
  Chemistry] {10.1146/annurev.physchem.55.091602.094449}, 55, 427

\bibitem[\protect\citeauthoryear{Mart{\'\i}nez, Castro, Rubio, Poblet  \&
  Alonso}{Mart{\'\i}nez et~al.}{2004}]{ref20}
Mart{\'\i}nez J.~I.,  Castro A.,  Rubio A.,  Poblet J.~M.,   Alonso J.~A.,
  2004, \mn@doi [Chemical Physics Letters]
  {https://doi.org/10.1016/j.cplett.2004.09.058}, 398, 292

\bibitem[\protect\citeauthoryear{Martinez, Castro, Rubio  \& Alonso}{Martinez
  et~al.}{2006}]{ref21}
Martinez J.~I.,  Castro A.,  Rubio A.,   Alonso J.~A.,  2006, \mn@doi [The
  Journal of Chemical Physics] {10.1063/1.2263732}, 125, 074311

\bibitem[\protect\citeauthoryear{May, Cartier  \& Castleman}{May
  et~al.}{1995}]{ref22}
May B.,  Cartier S.,   Castleman A.,  1995, \mn@doi [Chemical Physics Letters]
  {https://doi.org/10.1016/0009-2614(95)00739-Q}, 242, 265

\bibitem[\protect\citeauthoryear{Megha, Banerjee  \& Ghanty}{Megha
  et~al.}{2021}]{Megha2021}
Megha Banerjee A.,   Ghanty T.~K.,  2021, \mn@doi [Phys. Chem. Chem. Phys.]
  {10.1039/D0CP05756H}, 23, 5559

\bibitem[\protect\citeauthoryear{Okamoto}{Okamoto}{1998}]{ref10}
Okamoto H.,  1998, \mn@doi [Journal of Phase Equilibria]
  {10.1007/s12385-006-5014-8}, 19, 89

\bibitem[\protect\citeauthoryear{Patzer, Chang  \& S{\"u}lzle}{Patzer
  et~al.}{2014}]{ref39}
Patzer A. B.~C.,  Chang C.,   S{\"u}lzle D.,  2014, \mn@doi [Chemical Physics
  Letters] {https://doi.org/10.1016/j.cplett.2014.07.068}, 612, 39

\bibitem[\protect\citeauthoryear{Pfl\"uger, Fink, Weber, Bohnen  \&
  Crecelius}{Pfl\"uger et~al.}{1984}]{Pfluger1984}
Pfl\"uger J.,  Fink J.,  Weber W.,  Bohnen K.~P.,   Crecelius G.,  1984,
  \mn@doi [Phys. Rev. B] {10.1103/PhysRevB.30.1155}, 30, 1155

\bibitem[\protect\citeauthoryear{Rohmer, B{\'e}nard  \& Poblet}{Rohmer
  et~al.}{2000}]{ref14}
Rohmer M.-M.,  B{\'e}nard M.,   Poblet J.-M.,  2000, \mn@doi [Chemical Reviews]
  {10.1021/cr9803885}, 100, 495

\bibitem[\protect\citeauthoryear{Rubio, Balb\'as  \& Alonso}{Rubio
  et~al.}{1992}]{ref30}
Rubio A.,  Balb\'as L.~C.,   Alonso J.~A.,  1992, \mn@doi [Phys. Rev. B]
  {10.1103/PhysRevB.45.13657}, 45, 13657

\bibitem[\protect\citeauthoryear{Salazar-Villanueva, Hern{\'a}ndez~Tejeda, Pal,
  Rivas-Silva, Rodr{\'\i}guez~Mora  \& Ascencio}{Salazar-Villanueva
  et~al.}{2006}]{ref41}
Salazar-Villanueva M.,  Hern{\'a}ndez~Tejeda P.~H.,  Pal U.,  Rivas-Silva
  J.~F.,  Rodr{\'\i}guez~Mora J.~I.,   Ascencio J.~A.,  2006, \mn@doi [The
  Journal of Physical Chemistry A] {10.1021/jp061332e}, 110, 10274

\bibitem[\protect\citeauthoryear{Sholl \& Steckel}{Sholl \&
  Steckel}{2009}]{ref24}
Sholl D.~S.,  Steckel J.~A.,  2009, What is Density Functional Theory?.
John Wiley \& Sons, Ltd, pp 1--33,
  \mn@doi{https://doi.org/10.1002/9780470447710.ch1}

\bibitem[\protect\citeauthoryear{Sloan et~al.,}{Sloan et~al.}{2014}]{ref9}
Sloan G.~C.,  et~al., 2014, \mn@doi [The Astrophysical Journal]
  {10.1088/0004-637x/791/1/28}, 791, 28

\bibitem[\protect\citeauthoryear{Wang, Ding  \& Wang}{Wang
  et~al.}{1997}]{ref16}
Wang X.-B.,  Ding C.-F.,   Wang L.-S.,  1997, \mn@doi [The Journal of Physical
  Chemistry A] {10.1021/jp971838k}, 101, 7699

\bibitem[\protect\citeauthoryear{Wang, Wang, Wu  \& Cheng}{Wang
  et~al.}{1998}]{ref17}
Wang L.-S.,  Wang X.-B.,  Wu H.,   Cheng H.,  1998, \mn@doi [Journal of the
  American Chemical Society] {10.1021/ja9741990}, 120, 6556

\bibitem[\protect\citeauthoryear{Wei, Guo, Purnell, Buzza  \& Castleman}{Wei
  et~al.}{1992}]{ref12}
Wei S.,  Guo B.~C.,  Purnell J.,  Buzza S.,   Castleman A.~W.,  1992, \mn@doi
  [The Journal of Physical Chemistry] {10.1021/j100190a014}, 96, 4166

\bibitem[\protect\citeauthoryear{Wei, Zeng, You, Yan  \& Gong}{Wei
  et~al.}{2000}]{ref42}
Wei S.~H.,  Zeng Z.,  You J.~Q.,  Yan X.~H.,   Gong X.~G.,  2000, \mn@doi [The
  Journal of Chemical Physics] {10.1063/1.1319646}, 113, 11127

\bibitem[\protect\citeauthoryear{Wilhelm, Del~Ben  \& Hutter}{Wilhelm
  et~al.}{2016}]{ref38}
Wilhelm J.,  Del~Ben M.,   Hutter J.,  2016, \mn@doi [Journal of Chemical
  Theory and Computation] {10.1021/acs.jctc.6b00380}, 12, 3623

\bibitem[\protect\citeauthoryear{Yanai, Tew  \& Handy}{Yanai
  et~al.}{2004}]{ref36}
Yanai T.,  Tew D.~P.,   Handy N.~C.,  2004, \mn@doi [Chemical Physics Letters]
  {https://doi.org/10.1016/j.cplett.2004.06.011}, 393, 51

\bibitem[\protect\citeauthoryear{Yu, Huber  \& Froben}{Yu et~al.}{1995}]{ref15}
Yu H.,  Huber M.~G.,   Froben F.~W.,  1995, \mn@doi [Applied Surface Science]
  {https://doi.org/10.1016/0169-4332(94)00403-X}, 86, 74

\bibitem[\protect\citeauthoryear{Zhang, Jiang  \& Li}{Zhang
  et~al.}{2009}]{ref8}
Zhang K.,  Jiang B.~W.,   Li A.,  2009, \mn@doi [Monthly Notices of the Royal
  Astronomical Society] {10.1111/j.1365-2966.2009.14808.x}, 396, 1247

\bibitem[\protect\citeauthoryear{van Heijnsbergen, von Helden, Duncan, van Roij
   \& Meijer}{van Heijnsbergen et~al.}{1999}]{ref18}
van Heijnsbergen D.,  von Helden G.,  Duncan M.~A.,  van Roij A. J.~A.,
  Meijer G.,  1999, \mn@doi [Phys. Rev. Lett.] {10.1103/PhysRevLett.83.4983},
  83, 4983

\makeatother
\end{thebibliography}




\appendix

\section{Detailed photoabsorption spectra of small Ti$_{n}$C$_{x}$ clusters (n = 1--4; x = 1--4)} \label{apA}

\begin{figure}
\includegraphics[width=\columnwidth]{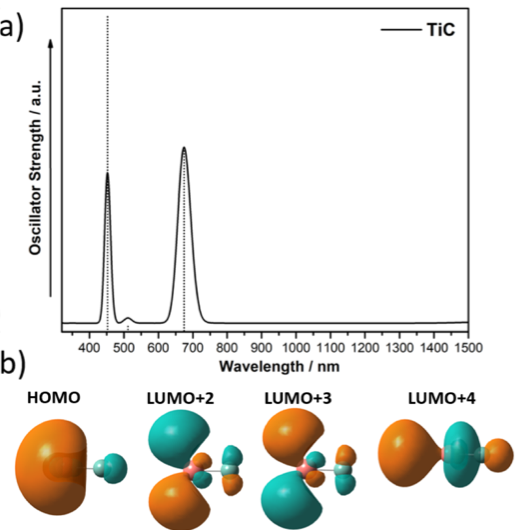}
\caption{Photoabsorption spectrum and frontier orbitals of TiC.}
\label{FigA1}
\end{figure}

\begin{figure}
\includegraphics[width=\columnwidth]{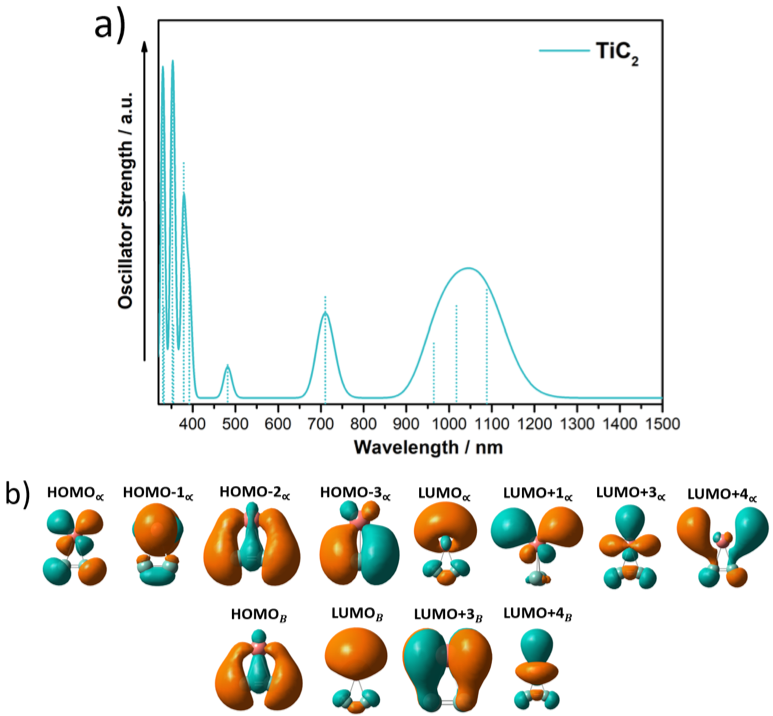}
\caption{Photoabsorption spectrum and frontier orbitals of TiC$_2$.}
\label{FigA2}
\end{figure}

\begin{figure}
\includegraphics[width=\columnwidth]{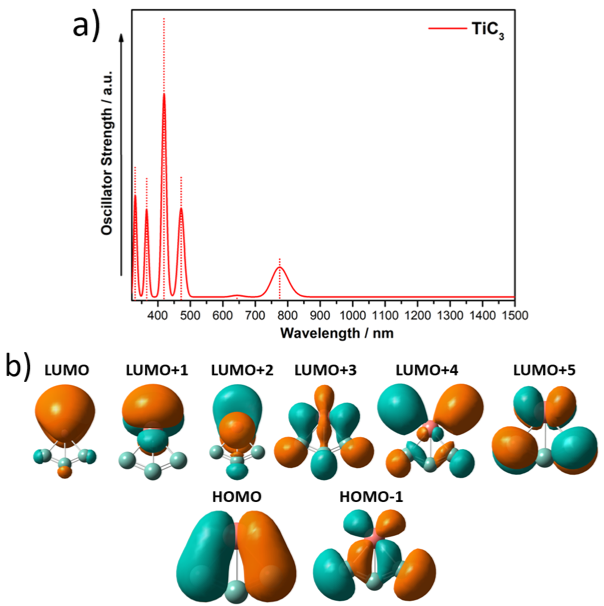}
\caption{Photoabsorption spectrum and frontier orbitals of TiC$_3$.}
\label{FigA3}
\end{figure}

\begin{figure}
\includegraphics[width=\columnwidth]{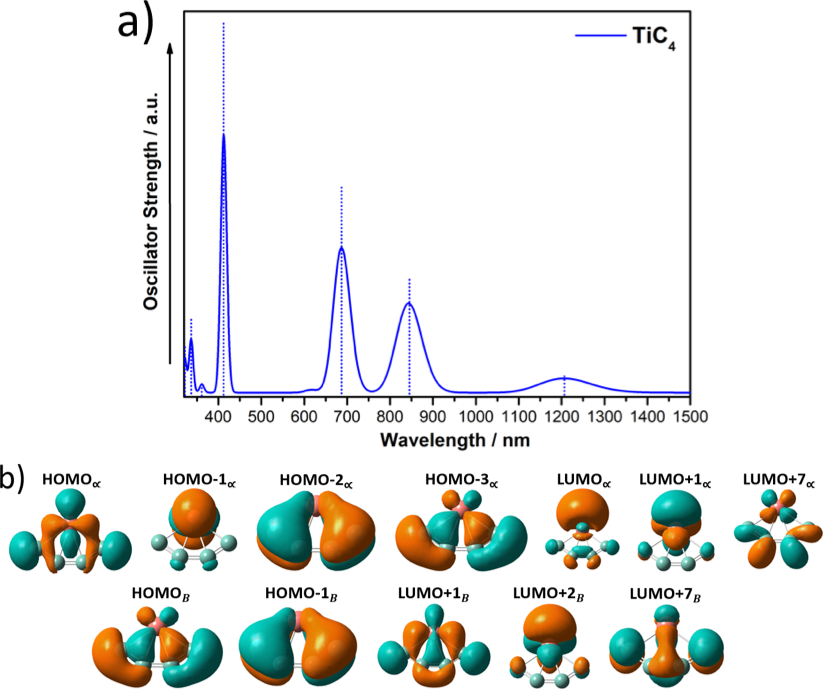}
\caption{Photoabsorption spectrum and frontier orbitals of TiC$_4$.}
\label{FigA4}
\end{figure}

\begin{figure}
\includegraphics[width=\columnwidth]{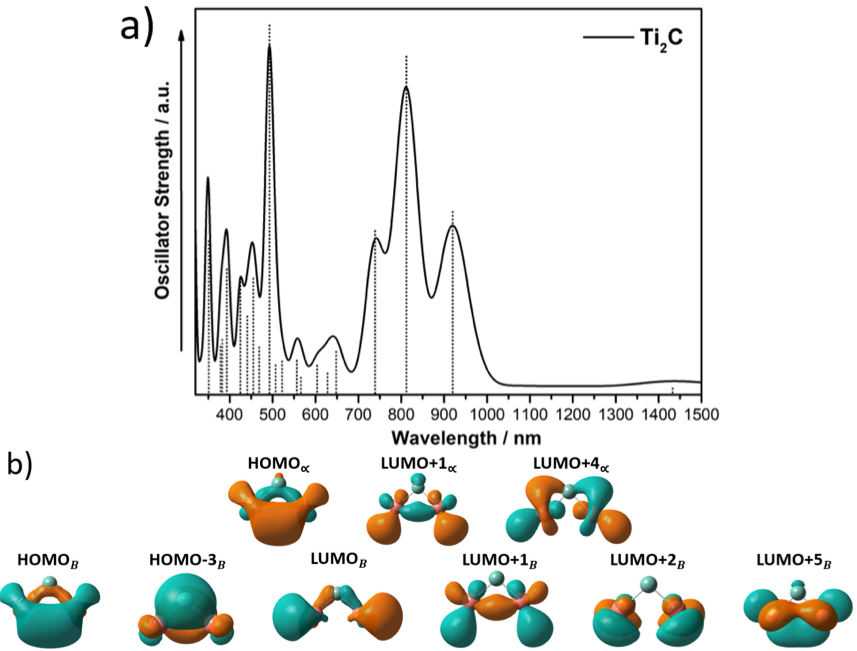}
\caption{Photoabsorption spectrum and frontier orbitals of Ti$_2$C.}
\label{FigA5}
\end{figure}

\begin{figure}
\includegraphics[width=\columnwidth]{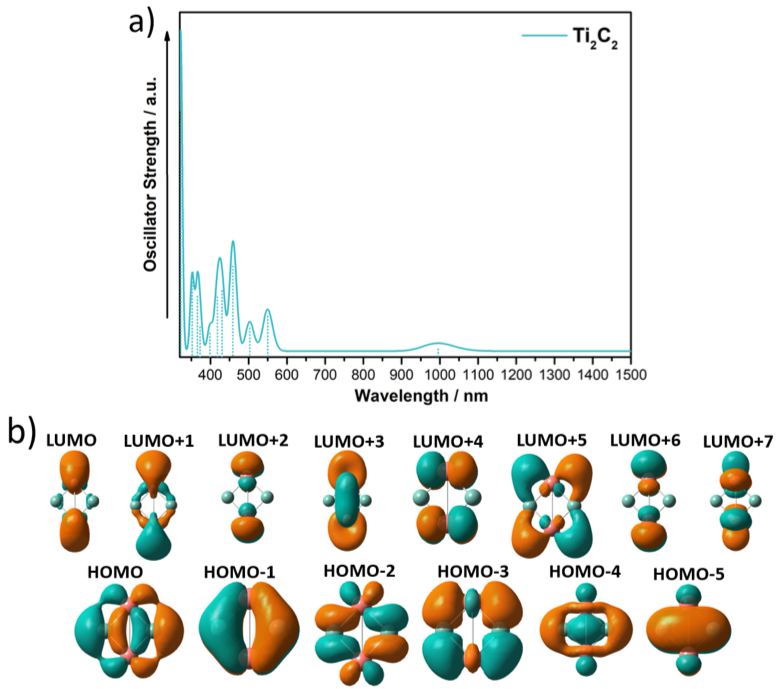}
\caption{Photoabsorption spectrum and frontier orbitals of Ti$_2$C$_2$.}
\label{FigA6}
\end{figure}

\begin{figure}
\includegraphics[width=\columnwidth]{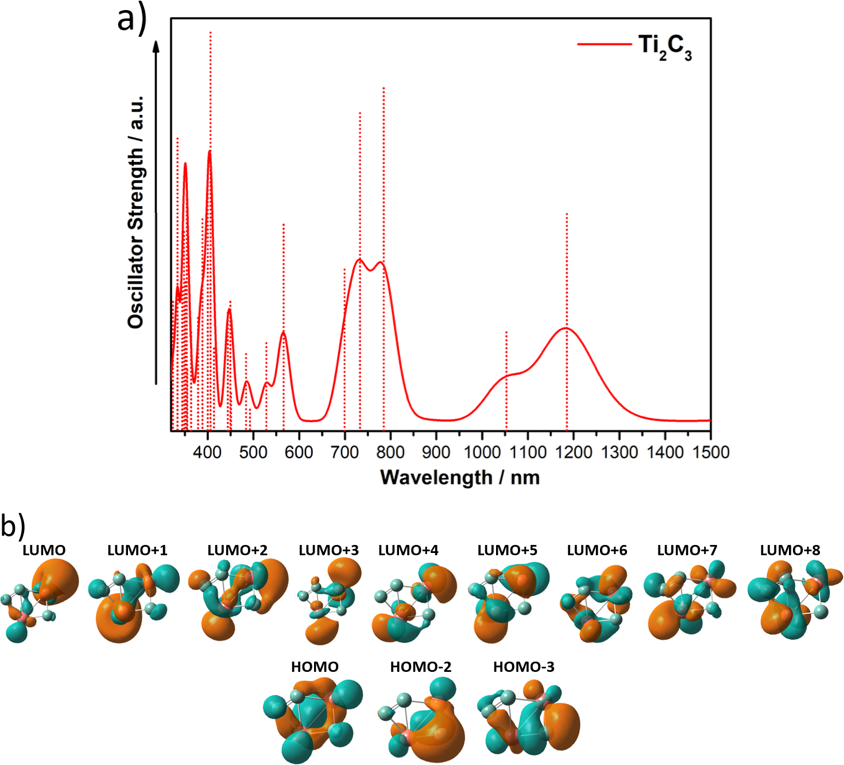}
\caption{Photoabsorption spectrum and frontier orbitals of Ti$_2$C$_3$.}
\label{FigA7}
\end{figure}

\begin{figure}
\includegraphics[width=\columnwidth]{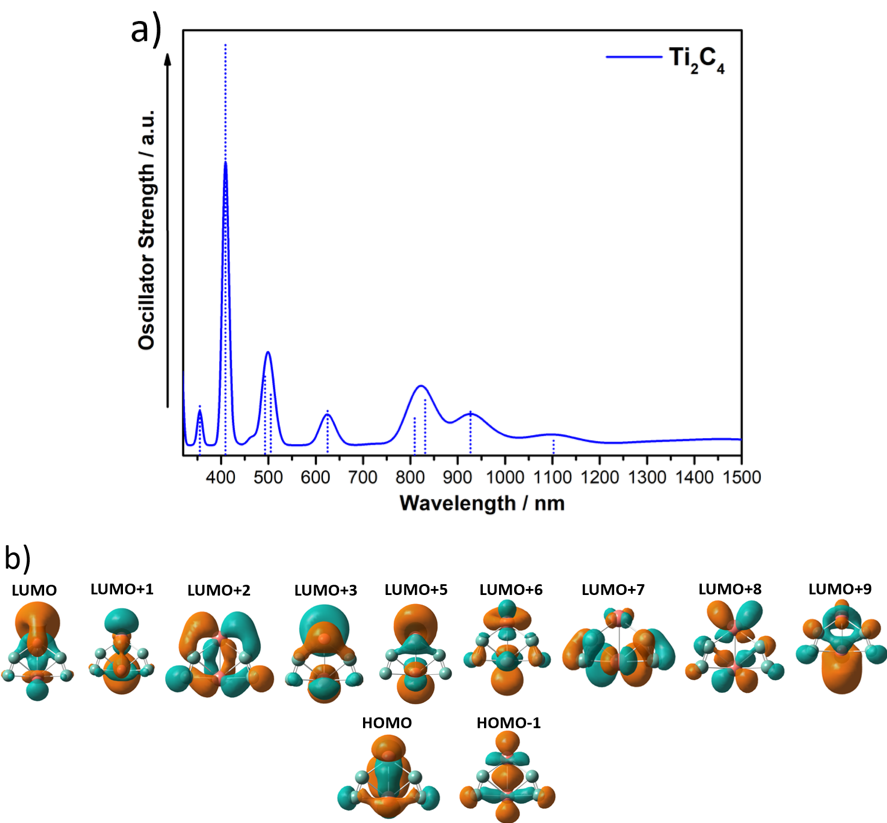}
\caption{Photoabsorption spectrum and frontier orbitals of Ti$_2$C$_4$.}
\label{FigA8}
\end{figure}

\begin{figure}
\includegraphics[width=\columnwidth]{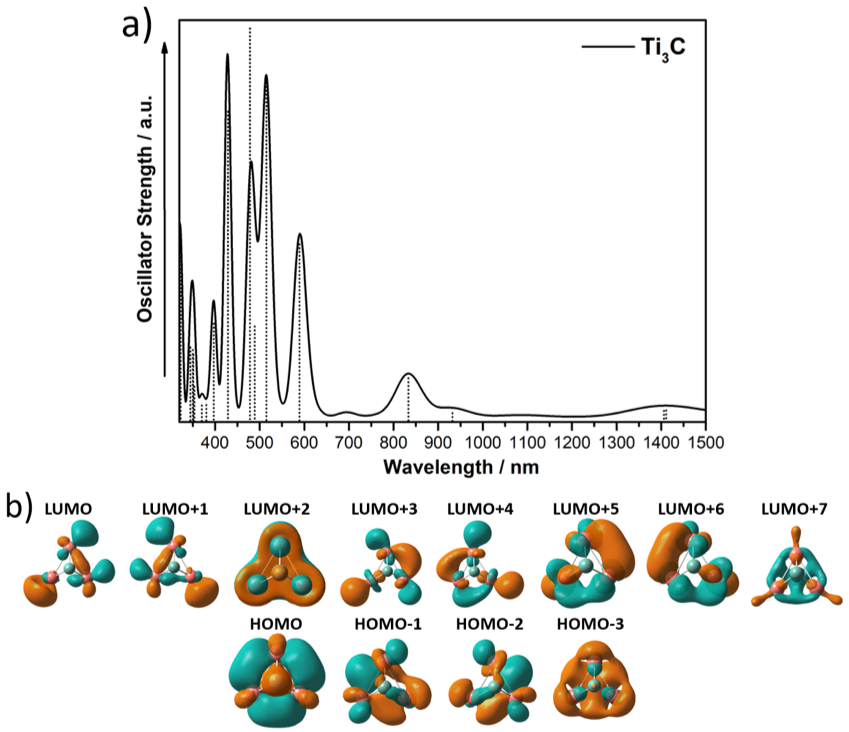}
\caption{Photoabsorption spectrum and frontier orbitals of Ti$_3$C.}
\label{FigA9}
\end{figure}

\begin{figure}
\includegraphics[width=\columnwidth]{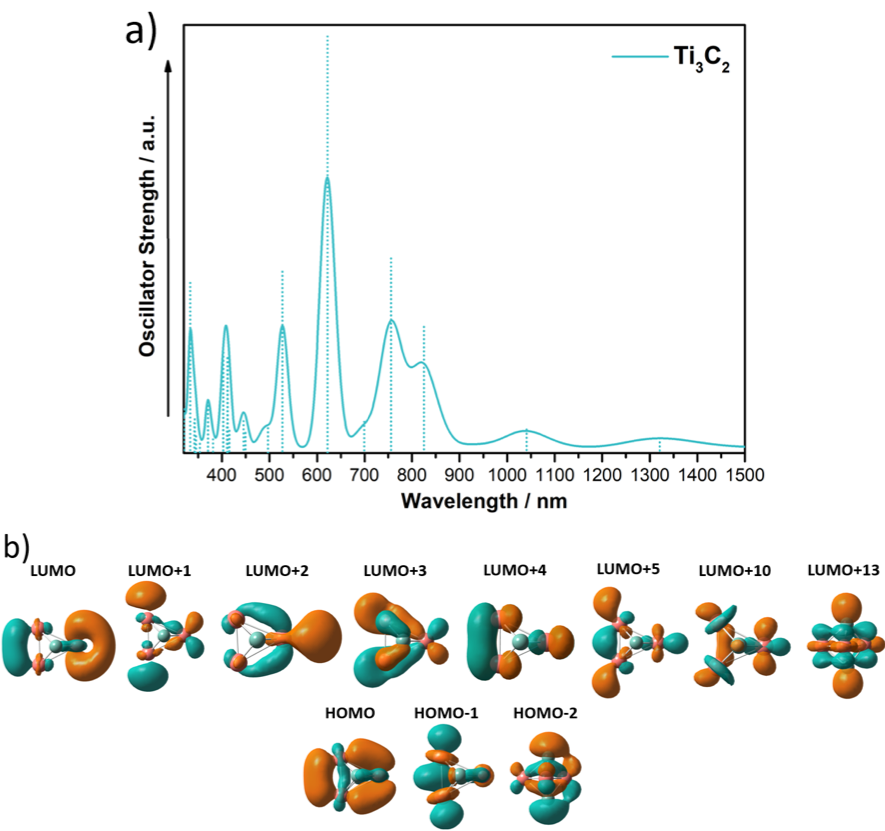}
\caption{Photoabsorption spectrum and frontier orbitals of Ti$_3$C$_2$.}
\label{FigA10}
\end{figure}

\begin{figure}
\includegraphics[width=\columnwidth]{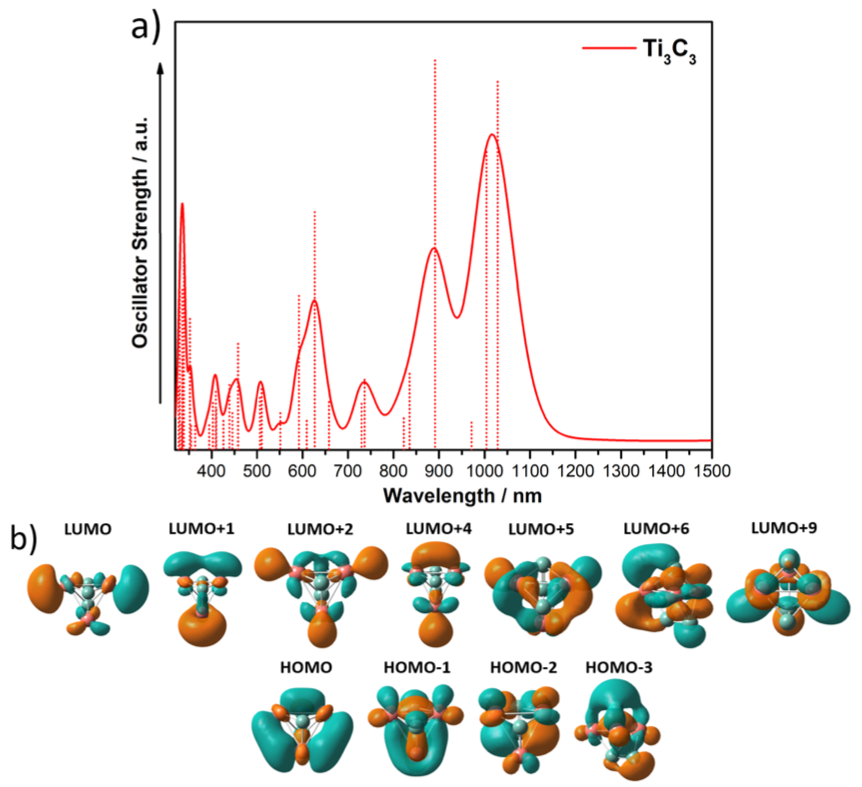}
\caption{Photoabsorption spectrum and frontier orbitals of Ti$_3$C$_3$.}
\label{FigA11}
\end{figure}

\begin{figure}
\includegraphics[width=\columnwidth]{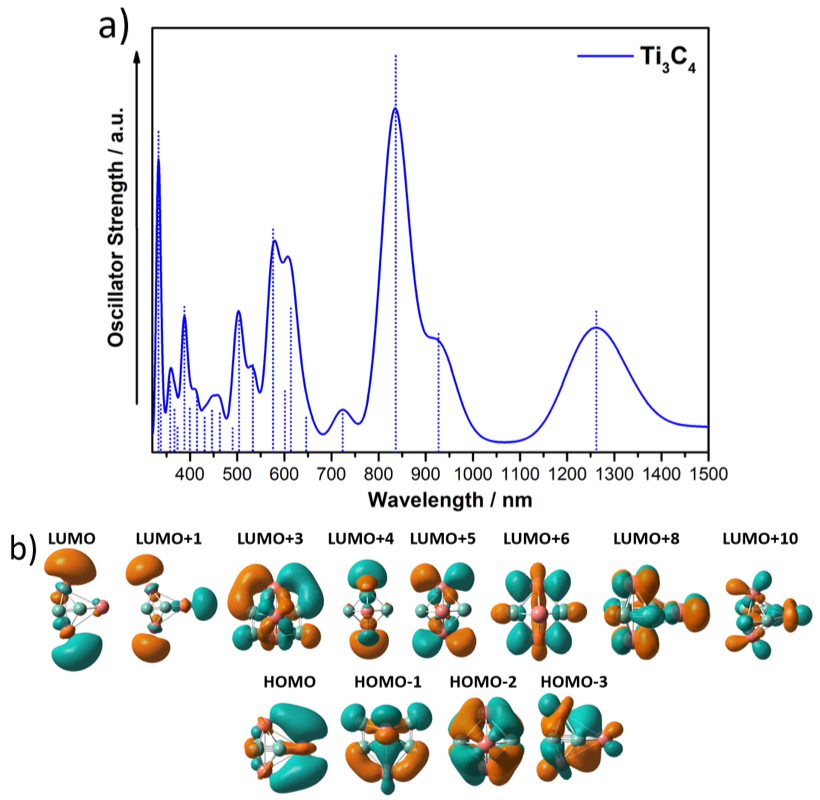}
\caption{Photoabsorption spectrum and frontier orbitals of Ti$_3$C$_4$.}
\label{FigA12}
\end{figure}

\begin{figure}
\includegraphics[width=\columnwidth]{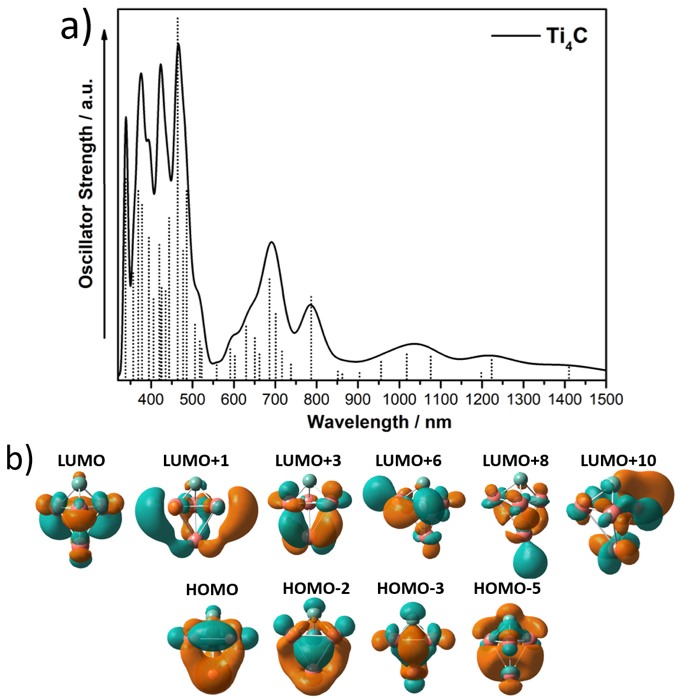}
\caption{Photoabsorption spectrum and frontier orbitals of Ti$_4$C.}
\label{FigA13}
\end{figure}

\begin{figure}
\includegraphics[width=\columnwidth]{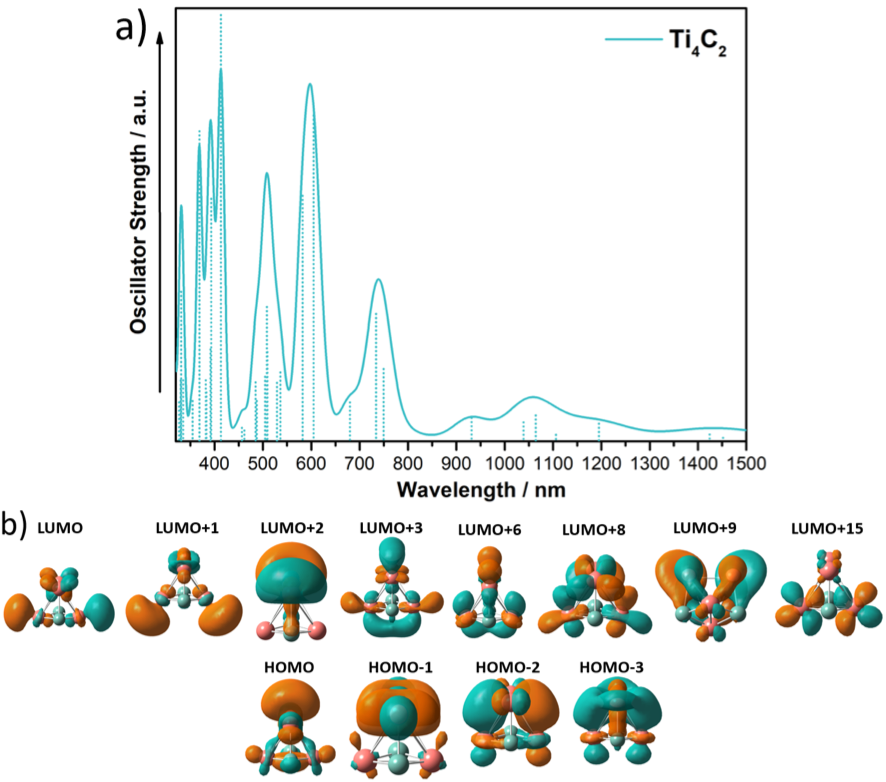}
\caption{Photoabsorption spectrum and frontier orbitals of Ti$_4$C$_2$.}
\label{FigA14}
\end{figure}

\begin{figure}
\includegraphics[width=\columnwidth]{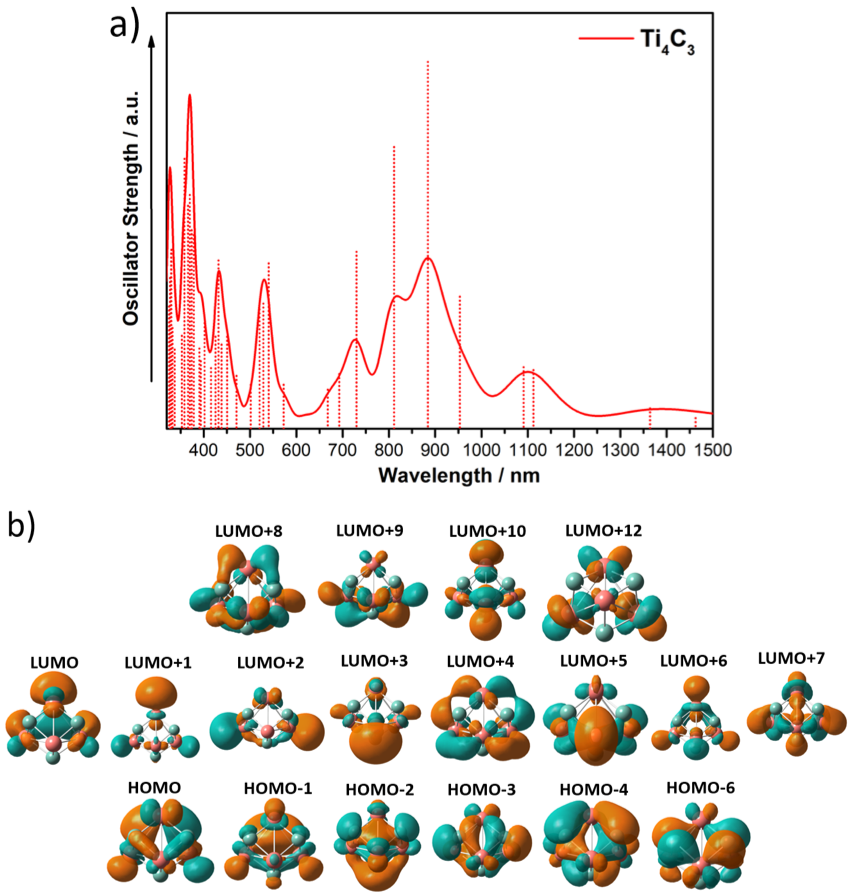}
\caption{Photoabsorption spectrum and frontier orbitals of Ti$_4$C$_3$.}
\label{FigA15}
\end{figure}

\begin{figure}
\includegraphics[width=\columnwidth]{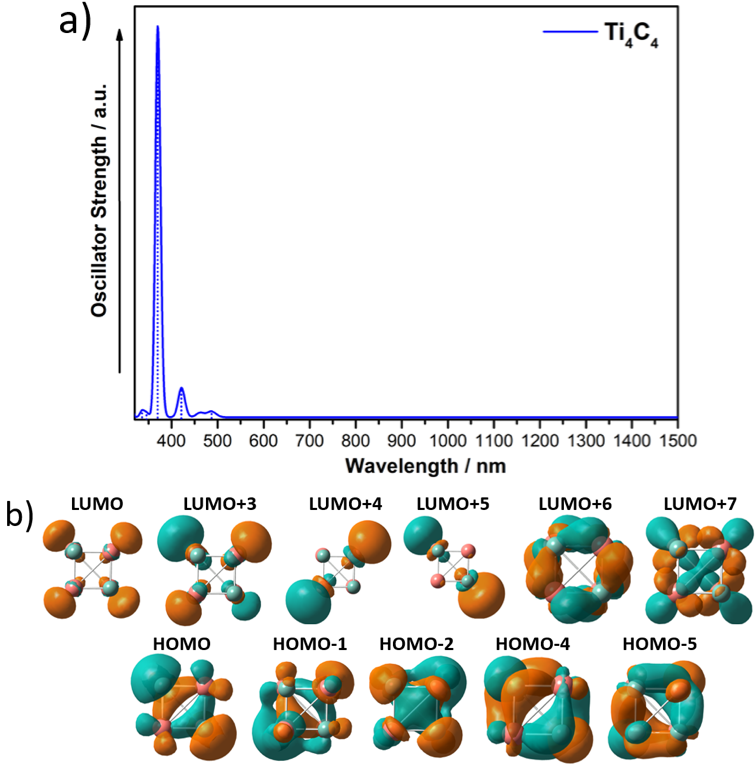}
\caption{Photoabsorption spectrum and frontier orbitals of Ti$_4$C$_4$.}
\label{FigA16}
\end{figure}

In this appendix we provide Figures~\ref{FigA1}--\ref{FigA16}, which show in panel (a) the calculated photoabsorption spectra of small Ti$_n$C$_x$ clusters (n = 1--4; x = 1--4), determined with TD-DFT at the CAM-B3LYP/6-31G** level of theory, in the range 320-1500 nm. The vertical bars indicate the positions (wavelengths) of the main excitations, and their oscillator strengths. The curves have been obtained by broadening the individual lines by gaussians of width 0.06 eV. Panel (b) gives the wavefunctions of the frontier molecular orbitals involved in the principal excitations. The Tables~\ref{TableA1}--\ref{TableA16} give the numerical values of the wavelengths and oscillator strengths for excitations with apreciable oscillator strength. The main single-particle-hole transitions contributing to each specific excitation are given in the column labelled Description. Those contributing transitions are characterized by the initial state (HOMO, HOMO-1, ...) and the final state (LUMO, LUMO+1, ...). In many cases, just one particle-hole transition is enough to characterize the excitation, but in other cases two, three or four different particle-hole transitions contribute to the excitation. Degenerate excitation lines are indicated by (*). $\alpha$ and $\beta$ in the orbitals represent the two spin orientations.

\begin{table}
\centering
\caption{Numerical values of the energies and wavelengths (in eV and nm, respectively), and oscillator strengths for excitations with appreciable oscillator strength for TiC. The main particle-hole transitions contributing to each excitation are given in the column labelled Description.}
	\label{TableA1}
	\begin{tabular}{cclc|lc}
		\hline
		Electronic \\ Transitions & Description & Energy & & & {\it f} \\
		\hline
		& & eV & nm & & \\
		\hline
		S$_0$$\rightarrow$S$_9$ & H$_{\alpha}$$\rightarrow$L+2$_{\alpha}$ (95\%) & 1.84  & 675.8 & & 0.091 \\
		S$_0$$\rightarrow$S$_{10}$ & H$_{\alpha}$$\rightarrow$L+3$_{\alpha}$ (95\%) & 1.84* & 675.8* & & 0.091 \\
		S$_0$$\rightarrow$S$_{18}$ & H-2$_{\alpha}$$\rightarrow$L$_{\alpha}$ (41\%) & 2.42  & 511.7 & & 0.0027 \\
		& H-1$_{\alpha}$$\rightarrow$L+1$_{\alpha}$ (41\%) & & & & \\
		& H-1$_{\beta}$$\rightarrow$L+3$_{\beta}$ (55\%) & & & & \\
		& H$_{\beta}$$\rightarrow$L+2$_{\beta}$ (55\%) & & & & \\
		S$_0$$\rightarrow$S$_{19}$ & H-2$_{\alpha}$$\rightarrow$L+1$_{\alpha}$ (41\%)& 2.42*  & 511.7* & & 0.0027 \\
		& H-1$_{\alpha}$$\rightarrow$L$_{\alpha}$ (42\%) & & & & \\
		& H-1$_{\beta}$$\rightarrow$L+2$_{\beta}$ (55\%) & & & & \\
		& H$_{\beta}$$\rightarrow$L+3$_{\beta}$ (55\%) & & & & \\
		S$_0$$\rightarrow$S$_{22}$ & H$_{\alpha}$$\rightarrow$L+4$_{\alpha}$ (96\%) & 2.74 & 452.5 & & 0.1559 \\
		\hline
	\end{tabular}
\end{table}

\begin{table}
\centering
\caption{Numerical values of the energies and wavelengths (in eV and nm, respectively), and oscillator strengths for excitations with appreciable oscillator strength for TiC$_2$. The main particle-hole transitions contributing to each excitation are given in the column labelled Description.}
	\label{TableA2}
	\begin{tabular}{cclc|lc}
		\hline
		Electronic \\ Transitions & Description & Energy & & & {\it f} \\
		\hline
		& & eV & nm & & \\
		\hline
		S$_0$$\rightarrow$S$_{5}$ & H-1$_{\alpha}$$\rightarrow$L+1$_{\alpha}$ (81\%) & 1.14  & 1087.6 & & 0.0064 \\
		S$_0$$\rightarrow$S$_{6}$ & H$_{\alpha}$$\rightarrow$L$_{\alpha}$ (75\%) & 1.22 & 1016.3 & & 0.0055 \\
		S$_0$$\rightarrow$S$_{7}$ & H-1$_{\alpha}$$\rightarrow$L$_{\alpha}$ (80\%) & 1.29 & 961.1 & & 0.0034 \\
		S$_0$$\rightarrow$S$_{8}$ & H-1$_{\alpha}$$\rightarrow$L+3$_{\alpha}$ (78\%) & 1.75  & 708.5 & & 0.0060 \\
		S$_0$$\rightarrow$S$_{9}$ & H-2$_{\alpha}$$\rightarrow$L$_{\alpha}$ (61\%) & 2.57 & 482.4 & & 0.0022 \\
				& H$_{\beta}$$\rightarrow$L$_{\beta}$ (60\%) & & & & \\
		S$_0$$\rightarrow$S$_{10}$ & H-2$_{\alpha}$$\rightarrow$L$_{\alpha}$ (65\%) & 3.16 & 392.4 & & 0.0076 \\
				& H$_{\beta}$$\rightarrow$L$_{\beta}$ (68\%) & & & & \\
		S$_0$$\rightarrow$S$_{12}$ & H-2$_{\alpha}$$\rightarrow$L+4$_{\alpha}$ (23\%) & 3.27  & 379.2 & & 0.0135 \\
				& H-3$_{\alpha}$$\rightarrow$L+2$_{\alpha}$ (15\%) & & & & \\
				& H$_{\beta}$$\rightarrow$L+3$_{\beta}$ (66\%) & & & & \\
		S$_0$$\rightarrow$S$_{15}$ & H$_{\alpha}$$\rightarrow$L+4$_{\alpha}$ (63\%) & 3.52 & 352.2 & & 0.019 \\
				& H$_{\beta}$$\rightarrow$L$_{\beta}$ (31\%) & & & & \\
				& H$_{\beta}$$\rightarrow$L+4$_{\beta}$ (24\%) & & & & \\
		\hline
	\end{tabular}
\end{table}

\begin{table}
\centering
\caption{Numerical values of the energies and wavelengths (in eV and nm, respectively), and oscillator strengths for excitations with appreciable oscillator strength for TiC$_3$. The main particle-hole transitions contributing to each excitation are given in the column labelled Description.}
	\label{TableA3}
	\begin{tabular}{cclc|lc}
		\hline
		Electronic \\ Transitions & Description & Energy & & & {\it f} \\
		\hline
		& & eV & nm & & \\
		\hline
		S$_0$$\rightarrow$S$_{2}$ & H$\rightarrow$L+1 (45\%) & 1.60  & 774.9 & & 0.0037 \\
				& H-1$\rightarrow$L (22\%) & & & & \\
				& H-1$\rightarrow$L+2 (21\%) & & & & \\
		S$_0$$\rightarrow$S$_{4}$ & H-1$\rightarrow$L (49\%) & 1.93 & 642.4 & & 0.0002 \\
				& H$\rightarrow$L+1 (35\%) & & & & \\
		S$_0$$\rightarrow$S$_{7}$ & H-1$\rightarrow$L+2 (65\%) & 2.63 & 471.4 & & 0.0110 \\
		S$_0$$\rightarrow$S$_{8}$ & H-1$\rightarrow$L+3 (86\%) & 2.96  & 418.9 & & 0.0252 \\
		S$_0$$\rightarrow$S$_{9}$ & H-1$\rightarrow$L+5 (47\%) & 3.39 & 365.7 & & 0.0109 \\
				& H$\rightarrow$L+4 (45\%) & & & & \\
		S$_0$$\rightarrow$S$_{11}$ & H-1$\rightarrow$L+4 (32\%) & 3.76 & 329.8 & & 0.0119 \\
				& H$\rightarrow$L+5 (58\%) & & & & \\
		\hline
	\end{tabular}
\end{table}

\begin{table}
\centering
\caption{Numerical values of the energies and wavelengths (in eV and nm, respectively), and oscillator strengths for excitations with appreciable oscillator strength for TiC$_4$. The main particle-hole transitions contributing to each excitation are given in the column labelled Description.}
	\label{TableA4}
	\begin{tabular}{cclc|lc}
		\hline
		Electronic \\ Transitions & Description & Energy & & & {\it f} \\
		\hline
		& & eV & nm & & \\
		\hline
		S$_0$$\rightarrow$S$_{3}$ & H$_{\alpha}$$\rightarrow$L+1$_{\alpha}$ (67\%) & 1.03  & 1203.7 & & 0.0015 \\
		S$_0$$\rightarrow$S$_{5}$ & H$_{\alpha}$$\rightarrow$L$_{\alpha}$ (95\%) & 1.47 & 843.4 & & 0.0085 \\
		S$_0$$\rightarrow$S$_{7}$ & H-1$_{\alpha}$$\rightarrow$L$_{\alpha}$ (95\%) & 1.80 & 688.8 & & 0.0152 \\
		S$_0$$\rightarrow$S$_{9}$ & H-3$_{\alpha}$$\rightarrow$L$_{\alpha}$ (27\%) & 3.01  & 411.9 & & 0.0271 \\
				& H$_{\alpha}$$\rightarrow$L+7$_{\alpha}$ (19\%) & & & & \\
				& H$_{\beta}$$\rightarrow$L+1$_{\beta}$ (81\%) & & & & \\
		S$_0$$\rightarrow$S$_{13}$ & H-2$_{\alpha}$$\rightarrow$L+1$_{\alpha}$ (85\%) & 3.69 & 336.0 & & 0.0056 \\
				& H-1$_{\beta}$$\rightarrow$L+2$_{\beta}$ (34\%) & & & & \\
				& H-1$_{\beta}$$\rightarrow$L+7$_{\beta}$ (22\%) & & & & \\
		\hline
	\end{tabular}
\end{table}

\begin{table}
\centering
\caption{Numerical values of the energies and wavelengths (in eV and nm, respectively), and oscillator strengths for excitations with appreciable oscillator strength for Ti$_2$C. The main particle-hole transitions contributing to each excitation are given in the column labelled Description.}
	\label{TableA5}
	\begin{tabular}{cclc|lc}
		\hline
		Electronic \\ Transitions & Description & Energy & & & {\it f} \\
		\hline
		& & eV & nm & & \\
		\hline
		S$_0$$\rightarrow$S$_{10}$ & H$_{\alpha}$$\rightarrow$L+4$_{\alpha}$ (70\%) & 1.35  & 918.4 & & 0.0474 \\
				& H$_{\beta}$$\rightarrow$L$_{\beta}$ (61\%) & & & & \\
		S$_0$$\rightarrow$S$_{12}$ & H$_{\beta}$$\rightarrow$L$_{\beta}$ (98\%) & 1.53 & 810.4 & & 0.0876 \\
		S$_0$$\rightarrow$S$_{16}$ & H$_{\alpha}$$\rightarrow$L+1$_{\alpha}$ (71\%) & 1.68 & 738.0 & & 0.0424 \\
				& H$_{\beta}$$\rightarrow$L+1$_{\beta}$ (86\%) & & & & \\
		S$_0$$\rightarrow$S$_{35}$ & H$_{\alpha}$$\rightarrow$L+1$_{\alpha}$ (51\%) & 2.52 & 492.0 & & 0.0957 \\
				& H$_{\beta}$$\rightarrow$L+2$_{\beta}$ (47\%) & & & & \\
		S$_0$$\rightarrow$S$_{39}$ & H-3$_{\beta}$$\rightarrow$L+1$_{\beta}$ (92\%) & 2.72 & 455.8 & & 0.0301 \\
		S$_0$$\rightarrow$S$_{53}$ & H-3$_{\beta}$$\rightarrow$L+5$_{\beta}$ (85\%) & 3.16 & 392.4 & & 0.0325 \\
		\hline
	\end{tabular}
\end{table}

\begin{table}
\centering
\caption{Numerical values of the energies and wavelengths (in eV and nm, respectively), and oscillator strengths for excitations with appreciable oscillator strength for Ti$_2$C$_2$. The main particle-hole transitions contributing to each excitation are given in the column labelled Description.}
	\label{TableA6}
	\begin{tabular}{cclc|lc}
		\hline
		Electronic \\ Transitions & Description & Energy & & & {\it f} \\
		\hline
		& & eV & nm & & \\
		\hline
		S$_0$$\rightarrow$S$_{8}$ & H-3$\rightarrow$L (53\%) & 2.26 & 548.6 & & 0.0153 \\
				& H-3$\rightarrow$L+3 (24\%) & & & & \\
				& H-1$\rightarrow$L+4 (17\%) & & & & \\
		S$_0$$\rightarrow$S$_{11}$ & H$\rightarrow$L+3 (45\%) & 2.47 & 502.0 & & 0.0108 \\
				& H-2$\rightarrow$L+1 (23\%) & & & & \\
		S$_0$$\rightarrow$S$_{15}$ & H-1$\rightarrow$L+4 (48\%) & 2.70 & 459.2 & & 0.0342 \\
				& H$\rightarrow$L+5 (28\%) & & & & \\
		S$_0$$\rightarrow$S$_{19}$ & H-5$\rightarrow$L (75\%) & 2.88  & 430.5 & & 0.0249 \\
		S$_0$$\rightarrow$S$_{21}$ & H-1$\rightarrow$L+2 (65\%) & 2.97 & 417.5 & & 0.0226 \\
		S$_0$$\rightarrow$S$_{29}$ & H-4$\rightarrow$L+1 (32\%) & 3.33 & 372.3 & & 0.0113 \\
				& H$\rightarrow$L+5 (28\%) & & & & \\
				& H-3$\rightarrow$L+3 (18\%) & & & & \\
		S$_0$$\rightarrow$S$_{36}$ & H-5$\rightarrow$L+6 (65\%) & 3.39 & 365.7 & & 0.0228 \\
		S$_0$$\rightarrow$S$_{37}$ & H-4$\rightarrow$L+7 (57\%) & 3.52 & 352.2 & & 0.0281 \\
		S$_0$$\rightarrow$S$_{41}$ & H-5$\rightarrow$L+6 (45\%) & 3.85 & 322.0 & & 0.1163 \\
				& H-3$\rightarrow$L+3 (23\%) & & & & \\
		\hline
	\end{tabular}
\end{table}

\begin{table}
\centering
\caption{Numerical values of the energies and wavelengths (in eV and nm, respectively), and oscillator strengths for excitations with appreciable oscillator strength for Ti$_2$C$_3$. The main particle-hole transitions contributing to each excitation are given in the column labelled Description.}
	\label{TableA7}
	\begin{tabular}{cclc|lc}
		\hline
		Electronic \\ Transitions & Description & Energy & & & {\it f} \\
		\hline
		& & eV & nm & & \\
		\hline
		S$_0$$\rightarrow$S$_{1}$ & H$\rightarrow$L+1 (48\%) & 0.69 & 1796.9 & & 0.0036 \\
				& H$\rightarrow$L (32\%) & & & & \\
		S$_0$$\rightarrow$S$_{2}$ & H$\rightarrow$L+5 (30\%) & 1.05 & 1180.8 & & 0.0121 \\
				& H$\rightarrow$L (27\%) & & & & \\
				& H$\rightarrow$L+1 (22\%) & & & & \\
		S$_0$$\rightarrow$S$_{3}$ & H$\rightarrow$L+2 (59\%) & 1.18 & 1050.7 & & 0.0055 \\
		S$_0$$\rightarrow$S$_{5}$ & H$\rightarrow$L+5 (33\%) & 1.58  & 784.7 & & 0.0191 \\
				& H$\rightarrow$L+6 (17\%) & & & & \\
		S$_0$$\rightarrow$S$_{6}$ & H$\rightarrow$L+6 (41\%) & 1.69 & 733.6 & & 0.0172 \\
				& H$\rightarrow$L+3 (13\%) & & & & \\
		S$_0$$\rightarrow$S$_{7}$ & H$\rightarrow$L+3 (36\%) & 1.77 & 700.5 & & 0.0090 \\
				& H$\rightarrow$L+4 (20\%) & & & & \\
		S$_0$$\rightarrow$S$_{8}$ & H$\rightarrow$L+7 (37\%) & 2.19 & 566.1 & & 0.0115 \\
				& H-1$\rightarrow$L (14\%) & & & & \\
		S$_0$$\rightarrow$S$_{18}$ & H-2$\rightarrow$L+4 (17\%) & 3.06 & 405.2 & & 0.0222 \\
				& H$\rightarrow$L+8 (14\%) & & & & \\
		S$_0$$\rightarrow$S$_{29}$ & H-3$\rightarrow$L+3 (27\%) & 3.71 & 334.2 & & 0.0163 \\
				& H-3$\rightarrow$L+6 (31\%) & & & & \\
		\hline
	\end{tabular}
\end{table}

\begin{table}
\centering
\caption{Numerical values of the energies and wavelengths (in eV and nm, respectively), and oscillator strengths for excitations with appreciable oscillator strength for Ti$_2$C$_4$. The main particle-hole transitions contributing to each excitation are given in the column labelled Description.}
	\label{TableA8}
	\begin{tabular}{cclc|lc}
		\hline
		Electronic \\ Transitions & Description & Energy & & & {\it f} \\
		\hline
		& & eV & nm & & \\
		\hline
		S$_0$$\rightarrow$S$_{1}$ & H$\rightarrow$L (78\%) & 0.50 & 2479.7 & & 0.0018 \\
		S$_0$$\rightarrow$S$_{6}$ & H$\rightarrow$L+3 (56\%) & 1.34 & 925.3 & & 0.0065 \\
		S$_0$$\rightarrow$S$_{7}$ & H-1$\rightarrow$L+3 (25\%) & 1.49 & 832.1 & & 0.0082 \\
				& H$\rightarrow$L+1 (20\%) & & & & \\
				& H-1$\rightarrow$L+1 (14\%) & & & & \\
				& H$\rightarrow$L (12\%) & & & & \\
		S$_0$$\rightarrow$S$_{8}$ & H-1$\rightarrow$L+2 (80\%) & 1.53 & 810.4 & & 0.0055 \\
		S$_0$$\rightarrow$S$_{11}$ & H-1$\rightarrow$L+5 (44\%) & 1.98  & 626.2 & & 0.0066 \\
				& H-1$\rightarrow$L+1 (22\%) & & & & \\
		S$_0$$\rightarrow$S$_{13}$ & H$\rightarrow$L+6 (51\%) & 2.46 & 504.0 & & 0.0090 \\
		S$_0$$\rightarrow$S$_{14}$ & H$\rightarrow$L+7 (50\%) & 2.52 & 492.0 & & 0.0117 \\
				& H$\rightarrow$L+8 (24\%) & & & & \\
		S$_0$$\rightarrow$S$_{17}$ & H-1$\rightarrow$L (27\%) & 3.03 & 409.2 & & 0.0608 \\
				& H-1$\rightarrow$L+5 (20\%) & & & & \\
				& H-1$\rightarrow$L+3 (18\%) & & & & \\
				& H-1$\rightarrow$L+6 (16\%) & & & & \\
		S$_0$$\rightarrow$S$_{18}$ & H$\rightarrow$L+9 (20\%) & 3.49 & 355.3 & & 0.0073 \\
				& H-1$\rightarrow$L+6 (17\%) & & & & \\
				& H$\rightarrow$L+5 (14\%) & & & & \\
		\hline
	\end{tabular}
\end{table}

\begin{table}
\centering
\caption{Numerical values of the energies and wavelengths (in eV and nm, respectively), and oscillator strengths for excitations with appreciable oscillator strength for Ti$_3$C. The main particle-hole transitions contributing to each excitation are given in the column labelled Description.}
	\label{TableA9}
	\begin{tabular}{cclc|lc}
		\hline
		Electronic \\ Transitions & Description & Energy & & & {\it f} \\
		\hline
		& & eV & nm & & \\
		\hline
		S$_0$$\rightarrow$S$_{20}$ & H$\rightarrow$L+3 (14\%) & 1.49 & 832.1 & & 0.0096 \\
				& H$\rightarrow$L+5 (7\%) & & & & \\
		S$_0$$\rightarrow$S$_{21}$ & H$\rightarrow$L+4 (15\%) & 1.49* & 832.1* & & 0.0098 \\
				& H$\rightarrow$L+6 (7\%) & & & & \\
		S$_0$$\rightarrow$S$_{30}$ & H-2$\rightarrow$L+4 (33\%) & 2.10 & 590.4 & & 0.0398 \\
				& H-3$\rightarrow$L+1 (25\%) & & & & \\
				& H-3$\rightarrow$L+3 (19\%) & & & & \\
		S$_0$$\rightarrow$S$_{32}$ & H-1$\rightarrow$L+7 (67\%) & 2.41 & 514.5 & & 0.0759 \\
		S$_0$$\rightarrow$S$_{33}$ & H-2$\rightarrow$L+7 (67\%) & 2.41*  & 514.5* & & 0.0754 \\
		S$_0$$\rightarrow$S$_{37}$ & H-3$\rightarrow$L+2 (58\%) & 2.59 & 478.7 & & 0.0881 \\
				& H$\rightarrow$L+7 (21\%) & & & & \\
		S$_0$$\rightarrow$S$_{40}$ & H-3$\rightarrow$L+5 (22\%) & 2.89 & 429.0 & & 0.0663 \\
				& H-3$\rightarrow$L (16\%) & & & & \\
		\hline
	\end{tabular}
\end{table}

\begin{table}
\centering
\caption{Numerical values of the energies and wavelengths (in eV and nm, respectively), and oscillator strengths for excitations with appreciable oscillator strength for Ti$_3$C$_2$. The main particle-hole transitions contributing to each excitation are given in the column labelled Description.}
	\label{TableA10}
	\begin{tabular}{cclc|lc}
		\hline
		Electronic \\ Transitions & Description & Energy & & & {\it f} \\
		\hline
		& & eV & nm & & \\
		\hline
		S$_0$$\rightarrow$S$_{3}$ & H-1$\rightarrow$L (57\%) & 0.94 & 1319.0 & & 0.0040 \\
				& H$\rightarrow$L (26\%) & & & & \\
		S$_0$$\rightarrow$S$_{4}$ & H$\rightarrow$L+1 (46\%) & 1.19 & 1041.9 & & 0.0063 \\
				& H$\rightarrow$L+3 (32\%) & & & & \\
		S$_0$$\rightarrow$S$_{7}$ & H-1$\rightarrow$L (28\%) & 1.50 & 826.6 & & 0.0339 \\
				& H$\rightarrow$L (23\%) & & & & \\
				& H$\rightarrow$L+5 (19\%) & & & & \\
		S$_0$$\rightarrow$S$_{10}$ & H$\rightarrow$L (39\%) & 1.64 & 756.0 & & 0.0523 \\
				& H$\rightarrow$L+5 (27\%) & & & & \\
		S$_0$$\rightarrow$S$_{15}$ & H-1$\rightarrow$L+1 (57\%) & 1.99  & 623.0 & & 0.1119 \\
		S$_0$$\rightarrow$S$_{17}$ & H-2$\rightarrow$L+4 (47\%) & 2.35 & 527.6 & & 0.0487 \\
				& H$\rightarrow$L+5 (14\%) & & & & \\
		S$_0$$\rightarrow$S$_{34}$ & H$\rightarrow$L+13 (27\%) & 3.01 & 411.9 & & 0.0255 \\
				& H-1$\rightarrow$L+13 (25\%) & & & & \\
				& H$\rightarrow$L+10 (11\%) & & & & \\
		S$_0$$\rightarrow$S$_{37}$ & H-2$\rightarrow$L+2 (41\%) & 3.08 & 402.6 & & 0.0270 \\
				& H-2$\rightarrow$L+5 (46\%) & & & & \\
		\hline
	\end{tabular}
\end{table}

\begin{table}
\centering
\caption{Numerical values of the energies and wavelengths (in eV and nm, respectively), and oscillator strengths for excitations with appreciable oscillator strength for Ti$_3$C$_3$. The main particle-hole transitions contributing to each excitation are given in the column labelled Description.}
	\label{TableA11}
	\begin{tabular}{cclc|lc}
		\hline
		Electronic \\ Transitions & Description & Energy & & & {\it f} \\
		\hline
		& & eV & nm & & \\
		\hline
		S$_0$$\rightarrow$S$_{1}$ & H$\rightarrow$L+1 (49\%) & 0.71 & 1746.3 & & 0.0019 \\
				& H-1$\rightarrow$L+1 (24\%) & & & & \\
		S$_0$$\rightarrow$S$_{4}$ & H$\rightarrow$L+1 (32\%) & 1.20 & 1033.2 & & 0.0430 \\
				& H-1$\rightarrow$L+1 (26\%) & & & & \\
		S$_0$$\rightarrow$S$_{5}$ & H-1$\rightarrow$L+5 (32\%) & 1.23 & 1008.0 & & 0.0348 \\
				& H$\rightarrow$L (30\%) & & & & \\
		S$_0$$\rightarrow$S$_{7}$ & H$\rightarrow$L (47\%) & 1.39 & 892.0 & & 0.0455 \\
				& H-1$\rightarrow$L+5 (21\%) & & & & \\
		S$_0$$\rightarrow$S$_{19}$ & H-1$\rightarrow$L+4 (18\%) & 1.98  & 626.2 & & 0.0278 \\
				& H$\rightarrow$L+6 (14\%) & & & & \\
				& H-1$\rightarrow$L+6 (13\%) & & & & \\
				& H$\rightarrow$L+2 (12\%) & & & & \\
				& H-1$\rightarrow$L+2 (9\%) & & & & \\
		S$_0$$\rightarrow$S$_{21}$ & H-2$\rightarrow$L+6 (42\%) & 2.09 & 593.2 & & 0.0181 \\
				& H-2$\rightarrow$L+2 (17\%) & & & & \\
		S$_0$$\rightarrow$S$_{30}$ & H-3$\rightarrow$L+1 (24\%) & 2.71 & 457.5 & & 0.0124 \\
				& H-1$\rightarrow$L+9 (21\%) & & & & \\
		\hline
	\end{tabular}
\end{table}

\begin{table}
\centering
\caption{Numerical values of the energies and wavelengths (in eV and nm, respectively), and oscillator strengths for excitations with appreciable oscillator strength for Ti$_3$C$_4$. The main particle-hole transitions contributing to each excitation are given in the column labelled Description.}
	\label{TableA12}
	\begin{tabular}{cclc|lc}
		\hline
		Electronic \\ Transitions & Description & Energy & & & {\it f} \\
		\hline
		& & eV & nm & & \\
		\hline
		S$_0$$\rightarrow$S$_{1}$ & H$\rightarrow$L (87\%) & 0.83 & 1493.8 & & 0.0034 \\
		S$_0$$\rightarrow$S$_{2}$ & H$\rightarrow$L+1 (41\%) & 0.98 & 1265.2 & & 0.0265 \\
				& H-1$\rightarrow$L+1 (31\%) & & & & \\
		S$_0$$\rightarrow$S$_{3}$ & H$\rightarrow$L+7 (55\%) & 1.34 & 925.3 & & 0.0223 \\
				& H$\rightarrow$L (20\%) & & & & \\
		S$_0$$\rightarrow$S$_{5}$ & H$\rightarrow$L (69\%) & 1.48 & 837.7 & & 0.0747 \\
		S$_0$$\rightarrow$S$_{16}$ & H$\rightarrow$L+6 (30\%) & 2.02  & 613.8 & & 0.0271 \\
				& H-2$\rightarrow$L+5 (15\%) & & & & \\
				& H$\rightarrow$L+10 (12\%) & & & & \\
		S$_0$$\rightarrow$S$_{20}$ & H$\rightarrow$L+10 (27\%) & 2.15 & 576.7 & & 0.0420 \\
				& H$\rightarrow$L+6 (26\%) & & & & \\
		S$_0$$\rightarrow$S$_{25}$ & H$\rightarrow$L+8 (44\%) & 2.33 & 532.1 & & 0.0164 \\
				& H-1$\rightarrow$L+3 (25\%) & & & & \\
		S$_0$$\rightarrow$S$_{26}$ & H-3$\rightarrow$L+6 (33\%) & 2.46 & 504.0 & & 0.0248 \\
				& H-3$\rightarrow$L+1 (15\%) & & & & \\
		S$_0$$\rightarrow$S$_{40}$ & H-1$\rightarrow$L+7 (41\%) & 3.20 & 387.5 & & 0.0274 \\
				& H-1$\rightarrow$L+4 (21\%) & & & & \\
		\hline
	\end{tabular}
\end{table}

\begin{table}
\centering
\caption{Numerical values of the energies and wavelengths (in eV and nm, respectively), and oscillator strengths for excitations with appreciable oscillator strength for Ti$_4$C. The main particle-hole transitions contributing to each excitation are given in the column labelled Description.}
	\label{TableA13}
	\begin{tabular}{cclc|lc}
		\hline
		Electronic \\ Transitions & Description & Energy & & & {\it f} \\
		\hline
		& & eV & nm & & \\
		\hline
		S$_0$$\rightarrow$S$_{26}$ & H-3$\rightarrow$L+1 (19\%) & 1.57 & 789.7 & & 0.0180 \\
				& H-3$\rightarrow$L+3 (17\%) & & & & \\
				& H-2$\rightarrow$L+1 (13\%) & & & & \\
		S$_0$$\rightarrow$S$_{33}$ & H-3$\rightarrow$L+6 (29\%) & 1.81 & 685.0 & & 0.0218 \\
		S$_0$$\rightarrow$S$_{51}$ & H$\rightarrow$L+8 (20\%) & 2.55 & 486.2 & & 0.0408 \\
				& H-5$\rightarrow$L (16\%) & & & & \\
				& H-2$\rightarrow$L+10 (13\%) & & & & \\
		S$_0$$\rightarrow$S$_{55}$ & H-5$\rightarrow$L (18\%) & 2.67 & 464.4 & & 0.0780 \\
                                   & H$\rightarrow$L+8 (15\%) & & & & \\
		S$_0$$\rightarrow$S$_{87}$ & H-5$\rightarrow$L+1 (25\%) & 3.36  & 369.0 & & 0.0408 \\
				& H-5$\rightarrow$L+3 (24\%) & & & & \\
		\hline
	\end{tabular}
\end{table}

\begin{table}
\centering
\caption{Numerical values of the energies and wavelengths (in eV and nm, respectively), and oscillator strengths for excitations with appreciable oscillator strength for Ti$_4$C$_2$. The main particle-hole transitions contributing to each excitation are given in the column labelled Description.}
	\label{TableA14}
	\begin{tabular}{cclc|lc}
		\hline
		Electronic \\ Transitions & Description & Energy & & & {\it f} \\
		\hline
		& & eV & nm & & \\
		\hline
		S$_0$$\rightarrow$S$_{16}$ & H$\rightarrow$L+1 (26\%) & 1.69 & 733.6 & & 0.0294 \\
				& H$\rightarrow$L+3 (19\%) & & & & \\
		S$_0$$\rightarrow$S$_{23}$ & H-3$\rightarrow$L (40\%) & 2.05 & 604.8 & & 0.0752 \\
				& H-2$\rightarrow$L+1 (19\%) & & & & \\
		S$_0$$\rightarrow$S$_{26}$ & H$\rightarrow$L+6 (54\%) & 2.13 & 582.1 & & 0.0567 \\
		S$_0$$\rightarrow$S$_{34}$ & H-3$\rightarrow$L+1 (33\%) & 2.44 & 508.1 & & 0.0310 \\
				& H-3$\rightarrow$L+3 (27\%) & & & & \\
		S$_0$$\rightarrow$S$_{52}$ & H-2$\rightarrow$L+8 (17\%) & 3.00  & 413.3 & & 0.0984 \\
				& H-3$\rightarrow$L+1 (14\%) & & & & \\
		S$_0$$\rightarrow$S$_{59}$ & H-3$\rightarrow$L+9 (31\%) & 3.16 & 392.4 & & 0.0560 \\
				& H-3$\rightarrow$L+2 (28\%) & & & & \\
		S$_0$$\rightarrow$S$_{68}$ & H$\rightarrow$L+15 (37\%) & 3.36 & 369.0 & & 0.0716 \\
		S$_0$$\rightarrow$S$_{85}$ & H-1$\rightarrow$L+15 (42\%) & 3.74 & 331.5 & & 0.0345 \\
		\hline
	\end{tabular}
\end{table}

\begin{table}
\centering
\caption{Numerical values of the energies and wavelengths (in eV and nm, respectively), and oscillator strengths for excitations with appreciable oscillator strength for Ti$_4$C$_3$. The main particle-hole transitions contributing to each excitation are given in the column labelled Description.}
	\label{TableA15}
	\begin{tabular}{cclc|lc}
		\hline
		Electronic \\ Transitions & Description & Energy & & & {\it f} \\
		\hline
		& & eV & nm & & \\
		\hline
		S$_0$$\rightarrow$S$_{5}$ & H$\rightarrow$L+2 (54\%) & 1.11 & 1117.0 & & 0.0060 \\
		S$_0$$\rightarrow$S$_{6}$ & H$\rightarrow$L+7 (41\%) & 1.14 & 1087.6 & & 0.0063 \\
		S$_0$$\rightarrow$S$_{7}$ & H$\rightarrow$L+8 (39\%) & 1.30 & 953.7 & & 0.0135 \\
				& H$\rightarrow$L+9 (20\%) & & & & \\
		S$_0$$\rightarrow$S$_{8}$ & H$\rightarrow$L+6 (50\%) & 1.40 & 885.6 & & 0.0375 \\
		S$_0$$\rightarrow$S$_{9}$ & H-1$\rightarrow$L (46\%) & 1.53 & 810.4 & & 0.0288 \\
		S$_0$$\rightarrow$S$_{11}$ & H-1$\rightarrow$L+5 (45\%) & 1.70 & 729.3 & & 0.0181 \\
		S$_0$$\rightarrow$S$_{18}$ & H-1$\rightarrow$L+1 (28\%) & 2.30 & 539.1 & & 0.0169 \\
				& H-2$\rightarrow$L+5 (12\%) & & & & \\
		S$_0$$\rightarrow$S$_{34}$ & H-3$\rightarrow$L+4 (13\%) & 2.87 & 432.0 & & 0.0172 \\
				& H-1$\rightarrow$L+3 (13\%) & & & & \\
				& H-4$\rightarrow$L+4 (10\%) & & & & \\
		S$_0$$\rightarrow$S$_{53}$ & H-4$\rightarrow$L+7 (24\%) & 3.35 & 370.1 & & 0.0239 \\
		S$_0$$\rightarrow$S$_{57}$ & H-1$\rightarrow$L+12 (28\%) & 3.46 & 358.3 & & 0.0276 \\
		S$_0$$\rightarrow$S$_{73}$ & H-6$\rightarrow$L+6 (12\%) & 3.76 & 329.8 & & 0.0183 \\
				& H-3$\rightarrow$L+10 (12\%) & & & & \\
				& H-6$\rightarrow$L+1 (11\%) & & & & \\
		\hline
	\end{tabular}
\end{table}

\begin{table}
\centering
\caption{Numerical values of the energies and wavelengths (in eV and nm, respectively), and oscillator strengths for excitations with appreciable oscillator strength for Ti$_4$C$_4$. The main particle-hole transitions contributing to each excitation are given in the column labelled Description.}
	\label{TableA16}
	\begin{tabular}{cclc|lc}
		\hline
		Electronic \\ Transitions & Description & Energy & & & {\it f} \\
		\hline
		& & eV & nm & & \\
		\hline
		S$_0$$\rightarrow$S$_{18}$ & H$\rightarrow$L (30\%) & 2.94 & 421.7 & & 0.0084 \\
				& H-4$\rightarrow$L+3 (19\%) & & & & \\
		S$_0$$\rightarrow$S$_{19}$ & H-1$\rightarrow$L (31\%) & 2.94* & 421.7* & & 0.0088 \\
				& H-5$\rightarrow$L+3 (17\%) & & & & \\
		S$_0$$\rightarrow$S$_{20}$ & H-2$\rightarrow$L (32\%) & 2.94* & 421.7* & & 0.0091 \\
				& H-5$\rightarrow$L+4 (14\%) & & & & \\
				& H-4$\rightarrow$L+5 (14\%) & & & & \\
		S$_0$$\rightarrow$S$_{38}$ & H-2$\rightarrow$L (22\%) & 3.35 & 370.1 & & 0.1208 \\
				& H-2$\rightarrow$L+7 (20\%) & & & & \\
		S$_0$$\rightarrow$S$_{39}$ & H$\rightarrow$L (24\%) & 3.35* & 370.1* & & 0.1208 \\
				& H-1$\rightarrow$L+6 (15\%) & & & & \\
		S$_0$$\rightarrow$S$_{40}$ & H-1$\rightarrow$L (22\%) & 3.35* & 370.1* & & 0.1208 \\
				& H$\rightarrow$L+6 (16\%) & & & & \\
		\hline
	\end{tabular}
\end{table}

\section{Detailed photoabsorption spectra of large Ti$_3$C$_8$, Ti$_4$C$_8$, Ti$_6$C$_{13}$, Ti$7$C$_{13}$, Ti$_8$C$_{12}$, Ti$_9$C$_{15}$, and Ti$_{13}$C$_{22}$ clusters} \label{apB}

\begin{figure}
\includegraphics[width=\columnwidth]{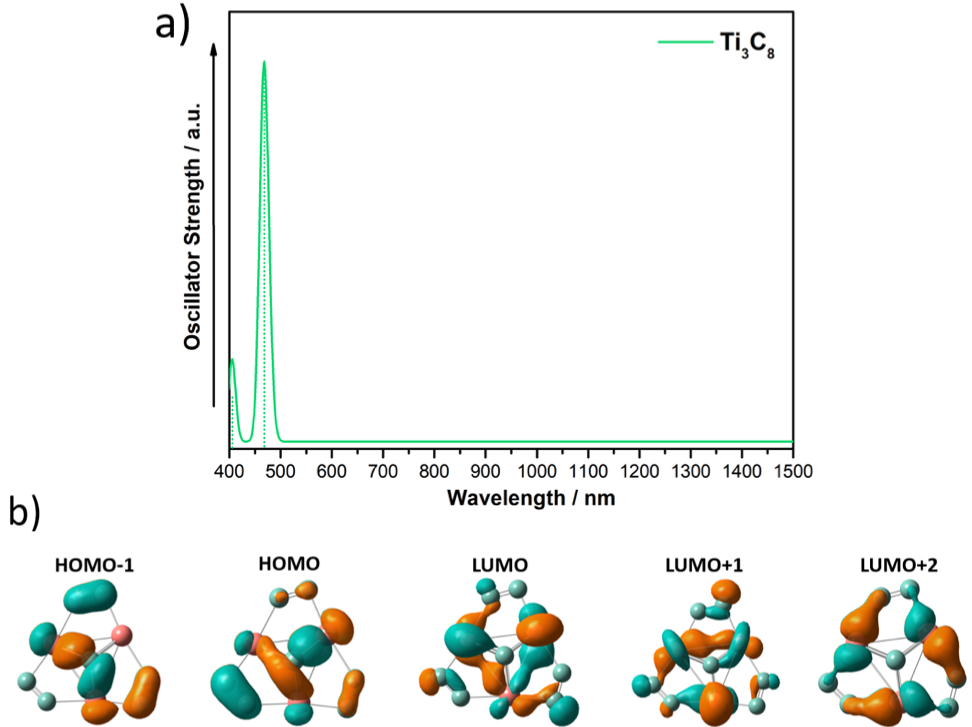}
\caption{Photoabsorption spectrum and frontier orbitals of Ti$_3$C$_8$.}
\label{FigB1}
\end{figure}

\begin{figure}
\includegraphics[width=\columnwidth]{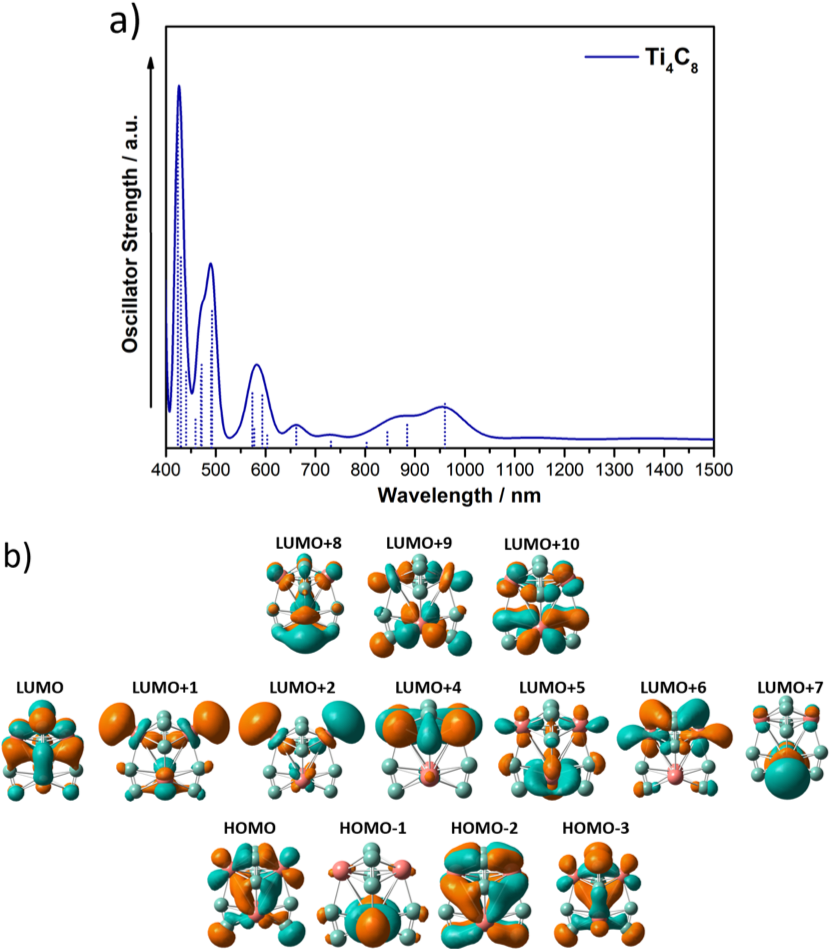}
\caption{Photoabsorption spectrum and frontier orbitals of Ti$_4$C$_8$.}
\label{FigB2}
\end{figure}

\begin{figure}
\includegraphics[width=\columnwidth]{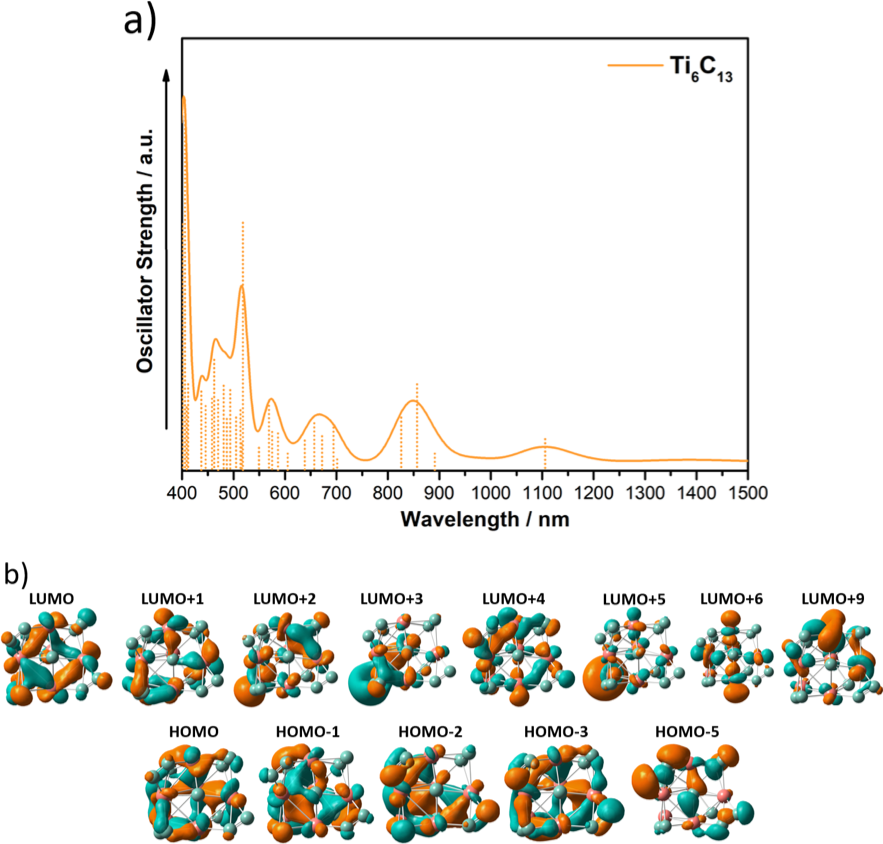}
\caption{Photoabsorption spectrum and frontier orbitals of Ti$_6$C$_{13}$.}
\label{FigB3}
\end{figure}

\begin{figure}
\includegraphics[width=\columnwidth]{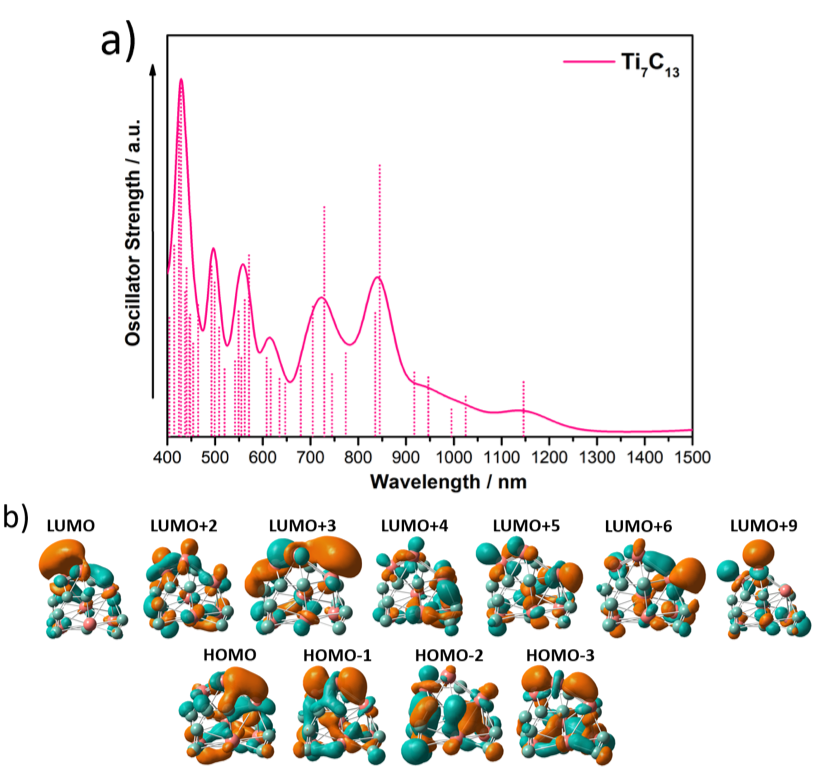}
\caption{Photoabsorption spectrum and frontier orbitals of Ti$_7$C$_{13}$.}
\label{FigB4}
\end{figure}

\begin{figure}
\includegraphics[width=\columnwidth]{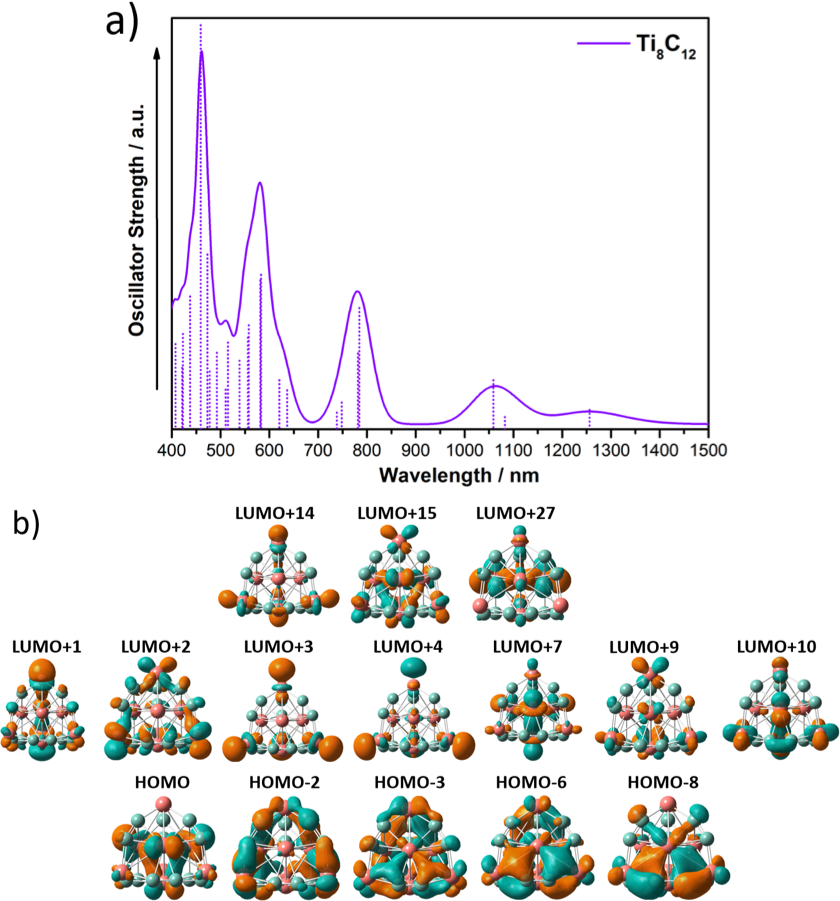}
\caption{Photoabsorption spectrum and frontier orbitals of Ti$_8$C$_{12}$.}
\label{FigB5}
\end{figure}

\begin{figure}
\includegraphics[width=\columnwidth]{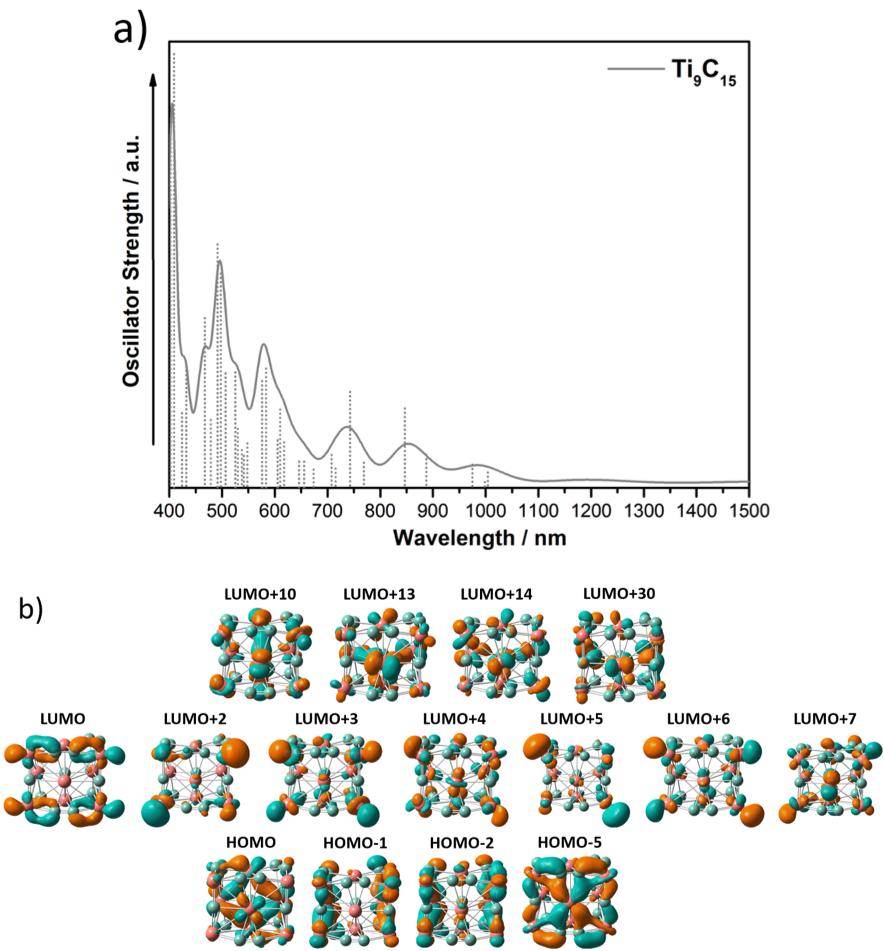}
\caption{Photoabsorption spectrum and frontier orbitals of Ti$_9$C$_{15}$.}
\label{FigB6}
\end{figure}

\begin{figure}
\includegraphics[width=\columnwidth]{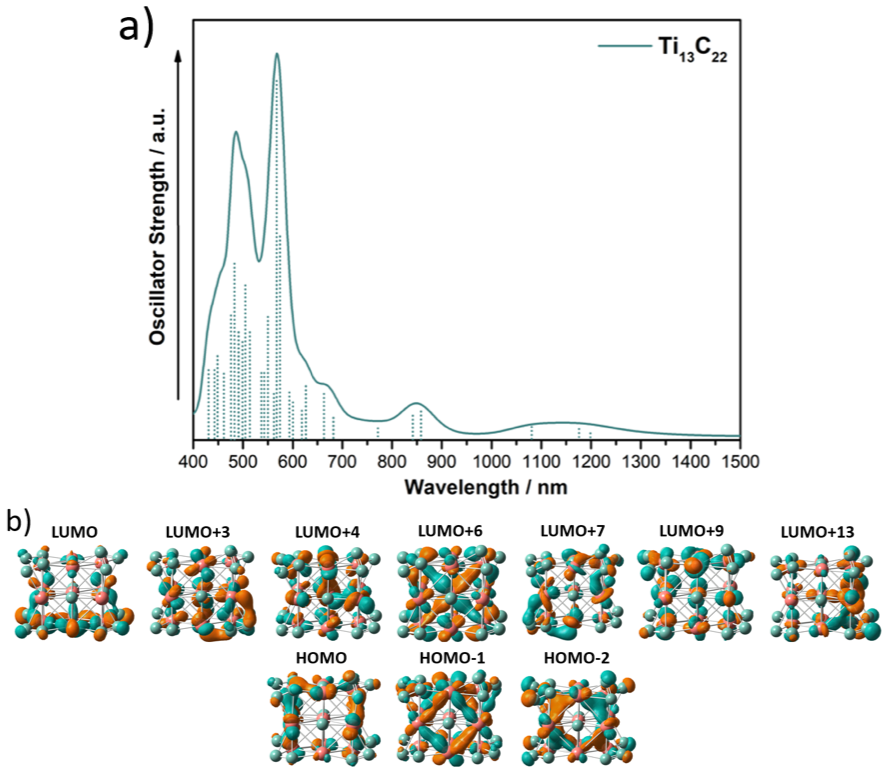}
\caption{Photoabsorption spectrum and frontier orbitals of Ti$_{13}$C$_{22}$.}
\label{FigB7}
\end{figure}

In this appendix we provide Figures~\ref{FigB1}--\ref{FigB7}, which show in panel (a) the calculated photoabsorption spectra of large Ti$_3$C$_8$, Ti$_4$C$_8$, Ti$_6$C$_{13}$, Ti$7$C$_{13}$, Ti$_8$C$_{12}$, Ti$_9$C$_{15}$, and Ti$_{13}$C$_{22}$, determined with TD-DFT at the CAM-B3LYP/6-31G** level of theory, in the range 400-1500 nm. The vertical bars indicate the positions (wavelengths) of the main excitations, and their oscillator strengths. The curves have been obtained by broadening the individual lines by gaussians of width 0.06 eV. Panel (b) gives the wavefunctions of the frontier molecular orbitals involved in the principal excitations. The Tables~\ref{TableB1}--\ref{TableB7} give the numerical values of the wavelengths and oscillator strengths for excitations with appreciable oscillator strength. The main particle-hole transitions contributing to each excitation are given in the column labelled Description. In many cases, just one particle-hole transition is enough to characterize the excitation, but in other cases two, three or four particle--hole transitions contribute to the excitation. Degenerate excitation lines are indicated by (*).

\begin{table}
\centering
\caption{Numerical values of the energies and wavelengths (in eV and nm, respectively), and oscillator strengths for excitations with appreciable oscillator strength for Ti$_3$C$_8$. The main particle-hole transitions contributing to each excitation are given in the column labelled Description.}
	\label{TableB1}
	\begin{tabular}{cclc|lc}
		\hline
		Electronic \\ Transitions & Description & Energy & & & {\it f} \\
		\hline
		& & eV & nm & & \\
		\hline
		S$_0$$\rightarrow$S$_{2}$ & H$\rightarrow$L+2 (71\%) & 2.65 & 467.9 & & 0.0023 \\
				& H$\rightarrow$L (20\%) & & & & \\
		S$_0$$\rightarrow$S$_{6}$ & H-1$\rightarrow$L+1 (20\%) & 3.06 & 405.2 & & 0.0001 \\
				& H-1$\rightarrow$L (13\%) & & & & \\
				& H$\rightarrow$L+1 (13\%) & & & & \\
		\hline
	\end{tabular}
\end{table}

\begin{table}
\centering
\caption{Numerical values of the energies and wavelengths (in eV and nm, respectively), and oscillator strengths for excitations with appreciable oscillator strength for Ti$_4$C$_8$. The main particle-hole transitions contributing to each excitation are given in the column labelled Description.}
	\label{TableB2}
	\begin{tabular}{cclc|lc}
		\hline
		Electronic \\ Transitions & Description & Energy & & & {\it f} \\
		\hline
		& & eV & nm & & \\
		\hline
		S$_0$$\rightarrow$S$_{4}$ & H$\rightarrow$L+2 (31\%) & 1.29 & 961.1 & & 0.0042 \\
				& H$\rightarrow$L+6 (18\%) & & & & \\
				& H-3$\rightarrow$L+1 (13\%) & & & & \\
		S$_0$$\rightarrow$S$_{5}$ & H$\rightarrow$L+1 (46\%) & 1.40 & 885.6 & & 0.0022 \\
				& H$\rightarrow$L+4 (39\%) & & & & \\
		S$_0$$\rightarrow$S$_{12}$ & H-3$\rightarrow$L (73\%) & 1.88 & 659.5 & & 0.0019 \\
		S$_0$$\rightarrow$S$_{16}$ & H-3$\rightarrow$L+4 (41\%) & 2.09 & 593.2 & & 0.0050 \\
				& H$\rightarrow$L+2 (35\%) & & & & \\
		S$_0$$\rightarrow$S$_{19}$ & H-1$\rightarrow$L (31\%) & 2.16 & 574.0 & & 0.0052 \\
				& H-1$\rightarrow$L+8 (29\%) & & & & \\
		S$_0$$\rightarrow$S$_{25}$ & H$\rightarrow$L+10 (39\%) & 2.52 & 492.0 & & 0.0130 \\
				& H-1$\rightarrow$L+7 (18\%) & & & & \\
		S$_0$$\rightarrow$S$_{28}$ & H-2$\rightarrow$L+6 (84\%) & 2.63 & 471.4 & & 0.0079 \\
		S$_0$$\rightarrow$S$_{33}$ & H-3$\rightarrow$L+6 (25\%) & 2.89 & 429.0 & & 0.0181 \\
				& H$\rightarrow$L+5 (18\%) & & & & \\
				& H$\rightarrow$L+6 (13\%) & & & & \\
		S$_0$$\rightarrow$S$_{34}$ & H$\rightarrow$L+9 (24\%) & 2.93 & 423.2 & & 0.0316 \\
		\hline
	\end{tabular}
\end{table}

\begin{table}
\centering
\caption{Numerical values of the energies and wavelengths (in eV and nm, respectively), and oscillator strengths for excitations with appreciable oscillator strength for Ti$_6$C$_{13}$. The main particle-hole transitions contributing to each excitation are given in the column labelled Description.}
	\label{TableB3}
	\begin{tabular}{cclc|lc}
		\hline
		Electronic \\ Transitions & Description & Energy & & & {\it f} \\
		\hline
		& & eV & nm & & \\
		\hline
		S$_0$$\rightarrow$S$_{3}$ & H$\rightarrow$L+1 (81\%) & 1.12 & 1107.0 & & 0.0022 \\
		S$_0$$\rightarrow$S$_{6}$ & H-1$\rightarrow$L+1 (75\%) & 1.45 & 855.1 & & 0.0060 \\
		S$_0$$\rightarrow$S$_{7}$ & H$\rightarrow$L+2 (32\%) & 1.50 & 826.6 & & 0.0037 \\
				& H-3$\rightarrow$L (24\%) & & & & \\
				& H$\rightarrow$L+3 (21\%) & & & & \\
		S$_0$$\rightarrow$S$_{10}$ & H-2$\rightarrow$L+1 (44\%) & 1.79 & 692.7 & & 0.0030 \\
				& H$\rightarrow$L+5 (13\%) & & & & \\
				& H$\rightarrow$L+6 (12\%) & & & & \\
		S$_0$$\rightarrow$S$_{11}$ & H-2$\rightarrow$L+1 (29\%) & 1.84 & 673.8 & & 0.0024 \\
				& H$\rightarrow$L+6 (13\%) & & & & \\
		S$_0$$\rightarrow$S$_{12}$ & H$\rightarrow$L+4 (40\%) & 1.89 & 656.0 & & 0.0033 \\
				& H-3$\rightarrow$L+1 (14\%) & & & & \\
				& H-2$\rightarrow$L+2 (12\%) & & & & \\
		S$_0$$\rightarrow$S$_{21}$ & H-1$\rightarrow$L+4 (14\%) & 2.39 & 518.8 & & 0.0172 \\
				& H-1$\rightarrow$L+5 (9\%) & & & & \\
				& H$\rightarrow$L+6 (8\%) & & & & \\
		S$_0$$\rightarrow$S$_{41}$ & H-5$\rightarrow$L (21\%) & 3.06 & 405.2 & & 0.0257 \\
				& H$\rightarrow$L+9 (8\%) & & & & \\
		\hline
	\end{tabular}
\end{table}

\begin{table}
\centering
\caption{Numerical values of the energies and wavelengths (in eV and nm, respectively), and oscillator strengths for excitations with appreciable oscillator strength for Ti$_7$C$_{13}$. The main particle-hole transitions contributing to each excitation are given in the column labelled Description.}
	\label{TableB4}
	\begin{tabular}{cclc|lc}
		\hline
		Electronic \\ Transitions & Description & Energy & & & {\it f} \\
		\hline
		& & eV & nm & & \\
		\hline
		S$_0$$\rightarrow$S$_{3}$ & H$\rightarrow$L (24\%) & 1.08 & 1148.0 & & 0.0034 \\
			& H$\rightarrow$L+2 (29\%) & & & & \\
		S$_0$$\rightarrow$S$_{11}$ & H-1$\rightarrow$L+2 (13\%) & 1.47 & 843.4 & & 0.0169 \\
			& H$\rightarrow$L+3 (8\%) & & & & \\
			& H$\rightarrow$L+5 (8\%) & & & & \\
			& H-3$\rightarrow$L (7\%) & & & & \\
		S$_0$$\rightarrow$S$_{19}$ & H-2$\rightarrow$L (7\%) & 1.70 & 729.3 & & 0.0143 \\
			& H-1$\rightarrow$L+9 (7\%) & & & & \\
			& H-1$\rightarrow$L+4 (5\%) & & & & \\
			& H-1$\rightarrow$L+6 (5\%) & & & & \\
		S$_0$$\rightarrow$S$_{35}$ & H-2$\rightarrow$L+5 (10\%) & 2.17 & 571.4 & & 0.0113 \\
				& H$\rightarrow$L+11 (5\%) & & & & \\
		\hline
	\end{tabular}
\end{table}

\begin{table}
\centering
\caption{Numerical values of the energies and wavelengths (in eV and nm, respectively), and oscillator strengths for excitations with appreciable oscillator strength for Ti$_8$C$_{12}$. The main particle-hole transitions contributing to each excitation are given in the column labelled Description.}
	\label{TableB5}
	\begin{tabular}{cclc|lc}
		\hline
		Electronic \\ Transitions & Description & Energy & & & {\it f} \\
		\hline
		& & eV & nm & & \\
		\hline
		S$_0$$\rightarrow$S$_{7}$ & H$\rightarrow$L+3 (80\%) & 0.99 & 1252.4 & & 0.0027 \\
		S$_0$$\rightarrow$S$_{10}$ & H$\rightarrow$L+4 (31\%) & 1.17 & 1059.7 & & 0.0068 \\
			& H$\rightarrow$L+27 (30\%) & & & & \\
		S$_0$$\rightarrow$S$_{12}$ & H$\rightarrow$L+10 (29\%) & 1.58 & 784.7 & & 0.0168 \\
			& H$\rightarrow$L+27 (25\%) & & & & \\
		S$_0$$\rightarrow$S$_{13}$ & H$\rightarrow$L+9 (63\%) & 1.59 & 779.8 & & 0.0105 \\
		S$_0$$\rightarrow$S$_{28}$ & H$\rightarrow$L+15 (32\%) & 2.12 & 584.8 & & 0.0214 \\
			& H-2$\rightarrow$L+2 (17\%) & & & & \\
		S$_0$$\rightarrow$S$_{30}$ & H$\rightarrow$L+14 (27\%) & 2.13 & 582.1 & & 0.0206 \\
			& H-3$\rightarrow$L+1 (24\%) & & & & \\
		S$_0$$\rightarrow$S$_{62}$ & H-8$\rightarrow$L+7 (17\%) & 2.70 & 459.2 & & 0.0560 \\
			& H-6$\rightarrow$L+1 (11\%) & & & & \\
		\hline
	\end{tabular}
\end{table}

\begin{table}
\centering
\caption{Numerical values of the energies and wavelengths (in eV and nm, respectively), and oscillator strengths for excitations with appreciable oscillator strength for Ti$_9$C$_{15}$. The main particle-hole transitions contributing to each excitation are given in the column labelled Description.}
	\label{TableB6}
	\begin{tabular}{cclc|lc}
		\hline
		Electronic \\ Transitions & Description & Energy & & & {\it f} \\
		\hline
		& & eV & nm & & \\
		\hline
		S$_0$$\rightarrow$S$_{9}$ & H$\rightarrow$L+13 (34\%) & 1.23 & 1008.0 & & 0.0016 \\
		S$_0$$\rightarrow$S$_{12}$ & H-1$\rightarrow$L+2 (28\%) & 1.27 & 976.3 & & 0.0024 \\
			& H-1$\rightarrow$L+3 (13\%) & & & & \\
		S$_0$$\rightarrow$S$_{14}$ & H$\rightarrow$L+7 (19\%) & 1.40 & 885.6 & & 0.0034 \\
			& H$\rightarrow$L+2 (19\%) & & & & \\
			& H$\rightarrow$L+14 (17\%) & & & & \\
		S$_0$$\rightarrow$S$_{15}$ & H-1$\rightarrow$L+3 (27\%) & 1.46 & 849.2 & & 0.0082 \\
			& H-1$\rightarrow$L+10 (10\%) & & & & \\
		S$_0$$\rightarrow$S$_{22}$ & H$\rightarrow$L+5 (19\%) & 1.67 & 742.4 & & 0.0099 \\
			& H-1$\rightarrow$L+13 (18\%) & & & & \\
		S$_0$$\rightarrow$S$_{40}$ & H$\rightarrow$L+30 (16\%) & 2.13 & 582.1 & & 0.0123 \\
			& H-5$\rightarrow$L+13 (13\%) & & & & \\
		S$_0$$\rightarrow$S$_{41}$ & H$\rightarrow$L+15 (9\%) & 2.15 & 576.7 & & 0.0110 \\
			& H-2$\rightarrow$L+4 (8\%) & & & & \\
			& H-2$\rightarrow$L+6 (8\%) & & & & \\
			& H-1$\rightarrow$L+10 (7\%) & & & & \\
		S$_0$$\rightarrow$S$_{60}$ & H-5$\rightarrow$L (36\%) & 2.52 & 492.0 & & 0.0251 \\
		\hline
	\end{tabular}
\end{table}

\begin{table}
\centering
\caption{Numerical values of the energies and wavelengths (in eV and nm, respectively), and oscillator strengths for excitations with appreciable oscillator strength for Ti$_13$C$_{22}$. The main particle-hole transitions contributing to each excitation are given in the column labelled Description.}
	\label{TableB7}
	\begin{tabular}{cclc|lc}
		\hline
		Electronic \\ Transitions & Description & Energy & & & {\it f} \\
		\hline
		& & eV & nm & & \\
		\hline
		S$_0$$\rightarrow$S$_{4}$ & H$\rightarrow$L+3 (24\%) & 1.03 & 1203.7 & & 0.0010 \\
				& H$\rightarrow$L+6 (17\%) & & & & \\
		S$_0$$\rightarrow$S$_{5}$ & H$\rightarrow$L (40\%) & 1.05 & 1180.8 & & 0.0016 \\
		S$_0$$\rightarrow$S$_{8}$ & H$\rightarrow$L+3 (23\%) & 1.15 & 1078.1 & & 0.0019 \\
			& H-1$\rightarrow$L (21\%) & & & & \\
			& H$\rightarrow$L (16\%) & & & & \\
		S$_0$$\rightarrow$S$_{16}$ & H-1$\rightarrow$L+3 (25\%) & 1.45 & 855.0 & & 0.0041 \\
			& H-2$\rightarrow$L+4 (12\%) & & & & \\
		S$_0$$\rightarrow$S$_{18}$ & H$\rightarrow$L+9 (15\%) & 1.47 & 843.4 & & 0.0035 \\
			& H-2$\rightarrow$L (12\%) & & & & \\
			& H$\rightarrow$L+7 (9\%) & & & & \\
		S$_0$$\rightarrow$S$_{55}$ & H-1$\rightarrow$L+13 (11\%) & 2.18 & 568.7 & & 0.0513 \\
			& H-1$\rightarrow$L+14 (8\%) & & & & \\
		\hline
	\end{tabular}
\end{table}


\bsp	
\label{lastpage}
\end{document}